\documentclass[a4paper,fleqn,usenatbib]{mnras}
\usepackage[T1]{fontenc}
\usepackage{ae,aecompl}

\usepackage{graphicx}	% Including figure files
\usepackage{amsmath}	% Advanced maths commands
\usepackage{amssymb}	% Extra maths symbols
\usepackage{epstopdf}
\usepackage{color}           % For color text: \color command
\usepackage{subfigure}
\DeclareTextSymbol{\degre}{T1}{6}

   \title[On the importance of astronomical refraction]{On the importance of astronomical refraction for modern solar astrometric measurements}

    \author[T. Corbard et al.]{T.~Corbard,$^1$\thanks{E-mail: corbard@oca.eu} R.~Ikhlef,$^{1,2}$ F.~Morand,$^1$ M.~Meftah,$^3$ and C.~Renaud$^1$\\
$^1$Universit\'e C\^ote d{\textquoteright}Azur, Observatoire de la C\^ote d{\textquoteright}Azur (OCA), CNRS, Laboratoire Lagrange, Bd. de l{\textquoteright}Observatoire CS 34229, \\
06304 Nice Cedex 4, France\\
$^2$ Centre de Recherche en Astronomie Astrophysique et G\'eophysique (CRAAG), BP:63 Bouzar\'eah 16340 Algiers, Algeria\\
$^3$ Universit\'e Paris Saclay, Universit\'e Pierre et Marie Curie, CNRS, LATMOS, 11 Boulevard d{\textquoteright}Alembert, 78280 Guyancourt, France }
          
\date{Accepted XXX. Received YYY; in original form ZZZ}

\pubyear{2018}

\begin{document}
\label{firstpage}
\pagerange{\pageref{firstpage}--\pageref{lastpage}}
\maketitle

% Abstract of the paper
\begin{abstract}

   In this work we study in details the influence of pure astronomical refraction on solar metrologic measurements made from ground-based full disk imagery and  provide
	the tools for correcting the measurements and estimating the associated uncertainties.
    For a given standard atmospheric model, we first  use both analytical and numerical methods in order to test the validity of the commonly or historically used approximations of the differential effect of refraction as a function of zenith distance. For a given refraction model, we provide the exact formulae for correcting solar radius measurements at any heliographic angle and for any zenith distance. Then, using solar images recorded in the near infrared between 2011 and 2016, we show that these corrections can be applied up to $70\degr$ using the usual approximate formulae and can be extended up to $80\degr$ of zenith distance provided that a standard atmospheric model and a full numerical integration of the refraction integral are used. 
	We also provide estimates of the absolute uncertainties associated with the differential 
	refraction corrections and show that approximate formulae can be used up to $80\degr$ of zenith distance for computing these uncertainties. For a given instrumental setup and the knowledge of the uncertainties associated with local weather records, this can be used to fix the maximum zenith distance one can observe depending on the required astrometric accuracy. 

\end{abstract}


%________________________________________________________________
%%%%%%%%%%%%%%%%%%%%%%%
\section{Introduction}
  \label{S-Introduction} 
%%%%%%%%%%%%%%%%%%%%%%%
	
	Ground based solar astrometric measurements have historically been made from transit instruments or astrolabes using the so-called equal altitudes method~\citep{Debarbat&Guinot1970}. Several instruments, derived from Danjon astrolabe, have been dedicated to solar diameter measurements, as DORaySol experiment \citep{Morand2010}. Observations consist in determining the transit times, through the same equal zenith distance circle, of the two solar  limbs which are the extremities of a vertical solar diameter.
As the two limbs are observed at equal zenith distances, influence of astronomical refraction is inherently reduced (e.g. Laclare et al. \citeyear{Laclare1996}). Only the small climatic conditions variations (temperature, pressure, relative humidity) between the two crossings, distant from a few minutes of time, can still play a role. 

Recent work in the field of solar metrology involve measurements from space using full disk solar images \citep{Dame1999, Kuhn2012, Meftah2015ApJ, Meftah2018} and planetary transits \citep{Hauchecorne2014ApJ, Emilio2015}. These measurements have however been made over a relatively short period of time and ground-based instruments were set-up in parallel to probe the long term variations of solar radius, their potential link with solar irradiance variations
	and their influence on Earth climate. 
	In order to test our ability to perform such measurements from ground on the long term, 
	similar techniques and instruments were used simultaneously from ground and space during the time of the PICARD space mission \citep{Meftah2014A&A, Meftah2015ApJ}. This  helped us to model and understand how the atmosphere affect ground based metrologic measurements. The main effect of atmospheric turbulence has been monitored with a dedicated instrument \citep{Ikhlef2016MNRAS} and calibrated corrections to radius measurement applied \citep{Meftah2018}. However, using full imagery from ground instead of the traditional astrolabe technique	also raise the question of the effect of refraction and how well we can correct from it.  Previous work used approximate formulae for refraction correction and a limit of $60\degr$ for the maximum observed zenith distance \citep{Meftah2014A&A, Meftah2015ApJ, Meftah2018}. With this conservative limit, no uncertainties were associated to the refraction corrections. The goal of this work is to use existing observations in order to test the validity of this observing limit and to associate uncertainties to refraction corrections as a function of the observing zenith distance.
	
	 The solar radius at any heliographic angle can be measured accurately at low zenith distances and we know from space measurements that the Sun is an almost perfect sphere with an oblateness less than $10^{-5}$ (e.g. Meftah et al. \citeyear{Meftah2015SoPh}). Solar astrometry can therefore be used to test our ability to correct the effect of field differential refraction as a function of zenith distance. In Section~\ref{sec:fund} we first recall the fundamental refraction integral which gives the link between the curvature of a light path in a given atmosphere and the apparent elevation of a celestial object. Some usual approximations of this integral, including the one that was historically used for solar astrometric measurements at Calern observatory are given in Section~\ref{sec:approx}. The field differential  refraction is then applied to the shape of the Sun. In Section~\ref{sec:shape} we first recall the usual  approximate formula which gives the refracted shape of the Sun as an ellipse  and then we establish the exact formula valid for any heliographic angle and zenith distance. We give in particular the inverse formula which allows to retrieve the true solar radius from the observed ones. In Section~\ref{sec:res} we use the full numerical integration of the refraction integral  and the exact equation for the shape of the Sun for a given standard atmosphere  in order to discuss the validity limits of the usual approximate formulae as a function of  zenith distance. Finally, in Section~\ref{sec:data} we use solar images recorded  in the near infrared in order to validate our correction of the differential refraction effects and  to test our ability to recover the true solar radius from the observed radii for zenith distances up to 80\degr. Our conclusions are given in Section~\ref{sec:conclu} where we outline the full procedure that we advocate for correcting solar radius measurements from differential refraction effects and estimating the associated uncertainties.

%%%%%%%%%%%%%%%%%%%%%%%%%%%%%%%%%%%%%%%%%%%%%%%%%%%%%%%%%%%%%%
\section{fundamental equations for astronomical refraction}
\label{sec:fund}
%%%%%%%%%%%%%%%%%%%%%%%%%%%%%%%%%%%%%%%%%%%%%%%%%%%%%%%%%%%%%%
The effect of atmospheric refraction is to change the true topocentric 
zenith angle $z^t$ of a celestial object to a lower observed one $z$. The refraction function $R(z)$ is defined by:
\begin{equation}\label{eq:gen}
z=z^t-R(z)
\end{equation}
Alternatively, we may take the true angles as argument and define the associated refraction function $\bar{R}$ by:
\begin{equation}\label{eq:gen_alt}
z=z^t-\bar{R}(z^t)  
\end{equation}
If the refraction function $R(z)$ is known, the associated function 
$\bar{R}(z^t)$ can easily be evaluated for any true zenith distance $z^t$
by solving the non linear equation $x-R(z^t-x) = 0$. 

From Snell's law of refraction applied to a spherical atmosphere, the curvature of a light path is linked to the local refractive index $n$ through the so-called refractive invariant: 
\begin{equation}\label{eq:invarient}
n \ r \sin(\xi)=\mathrm{constant}
\end{equation}
where  $\xi$ is the local zenith distance i.e. the angle between the light ray and the radius vector $r$ from 
Earth center. From this, the differential refraction along the light ray is obtained by:
\begin{equation}\label{eq:diff}
\mathrm{d}R=- \tan \xi \ \frac{\mathrm{d}n}{n}
\end{equation}
In order to find the total amount of refraction at observer position, 
we can integrate  along the full ray path from $n=n_{\mbox{obs}}$ and $\xi=z$ at observer position  up to 
$n=1$ outside the atmosphere. 
\begin{equation}\label{eq:int}
R= \int_1^{n_{\mathrm{obs}}} \tan \xi \ \frac{\mathrm{d}n}{n}
\end{equation}
This can be done either by direct numerical integration of Eq.~(\ref{eq:int}) after an 
appropriate change of variable~\citep{Auer&Standish2000} or by using a full ray-tracing procedure solving the system
of coupled differential equations provided by Eq.~(\ref{eq:diff}) and the differentiation of Eq.~(\ref{eq:invarient}) \citep{vanderWerf2003, vanderWerf2008}. This, in principle, requires a model of the full atmosphere i.e. temperature, pressure, density etc. at any point through the light path. In the next section we recall why this is in fact not 
needed if we avoid areas close to the horizon  and give some usual approximations of the refraction integral.

%%%%%%%%%%%%%%%%%%%%%%%%%%%%%%%%%%%%%%%%%%%%%%%%%%%
\section{Approximation to the refraction integral} 
\label{sec:approx}
%%%%%%%%%%%%%%%%%%%%%%%%%%%%%%%%%%%%%%%%%%%%%%%%%%
For zenith distance up to $70\degr$, the refraction integral can be evaluated with good accuracy without any hypothesis about the structure of the atmosphere: it depends only on temperature and pressure at the observer (Oriani's theorem, see also: \cite{Ball1908,Young2004, Young2006Obs}). This justifies that, over time, a large number of nearly equivalent 
approximate formulae have  been derived that do not require the full knowledge of the structure of the real atmosphere. A development of the refraction integral into  semi-convergent series of odd power of $\tan(z)$ is what is commonly found in textbooks~\citep{Ball1908, Smart1965, Woolard&Clemence1966, Danjon1980}.  
 An example of this will be given in Section~\ref{sec:calern}. In fact the first two  terms of such expansion (up to $\tan^3$) corresponds to what is known as Laplace formula of which \cite{Fletcher1931} said that \emph{ no reasonable theory differs by more than a few thousandths, hundredths, tenths of a second at $z= 60\degr$, $70\degr$, $75\degr$ respectively}. 

For large zenith distance, $\tan(z)$ power series will diverge at the horizon and are not appropriate. Closed formula 
valid at low zenith distance and that are finite at the horizon can however still be found (see e.g.~\cite{Wittmann1997AN}). Assuming an exponential law for the variation of air density with height, it's possible for instance to derive a formula involving 
the error function~\citep{Fletcher1931, Danjon1980}. Another example
 is Cassini's exact formula for an homogeneous atmosphere model. While physically un-realistic, the model of Cassini, thanks to Oriani's theorem, gives also excellent results up to at least $70\degr$ of zenith distance while remaining finite down to the horizon~\citep{Young2004}.
 For large zenith distances however, \cite{Young2004} have shown that the lowest layers of the atmosphere and especially the lapse rate at observer becomes progressively dominant as one observe closer to the horizon. This therefore should be included  in atmospheric models and we can not avoid anymore the full numerical evaluation of the refraction integral.
 
 In the following sub-sections we present first in details the refraction model as it was used for reducing solar astrolabe data at Calern observatory, then we give the full error function model from which the Calern model was actually derived and finally we recall Cassini's formula. In Section~\ref{sec:res},  these three approximations will then be compared to full numerical integration of the refraction integral using a standard atmosphere model.  

 %%%%%%%%%%%%%%%%%%%%%%%%%%%%%%%%%%%%%%%%%%%%%%%%%%%%%%%%%%%%%%%%%%%%%%%%%%%%%
\subsection{Refraction model used at Calern observatory for solar metrology}
\label{sec:calern}
%%%%%%%%%%%%%%%%%%%%%%%%%%%%%%%%%%%%%%%%%%%%%%%%%%%%%%%%%%%%%%%%%%%%%%%%%%

The refraction model that was used for the reduction of astrolabe measurements
at Calern observatory is a truncation of the expansion in odd power of $\tan(z)$ 
\citep{Danjon1980}. For an observer at geodetic latitude $\varphi$ and altitude $h$ 
above the reference ellipsoid,
the refraction $R$ is obtained as a function of the observed zenith angle, the wavelength 
($\lambda$) and local atmospheric conditions i.e. pressure ($P$),  absolute 
temperature ($T$), and relative humidity ($f_h\in[0,1]$)  by:

\begin{eqnarray}
\label{eq:Danjon}
R(z,\lambda,P,T,f_h,h,\varphi)&=\alpha(1-\beta) \tan(z)-\alpha(\beta-{\alpha \over 2}) \tan^3(z)   \nonumber\\
                    &   +3\alpha\left(\beta-{\frac{\alpha}{2}}\right)^2 \tan^5(z)  
\end{eqnarray}

where 
\begin{equation}
\label{eq:alpha1}
\alpha(T,P,f_h,\lambda)=n_{\mathrm{obs}}-1 
\end{equation}
is the air refractivity for local atmospheric conditions and the given wavelength, and
\begin{equation}
\label{eq:beta}
\beta(T,h,\varphi)=\ell(T)/r_c(\varphi,h)
\end{equation}
 is the ratio between the height $\ell$ of the homogeneous atmosphere and the Earth radius of 
curvature $r_c$ at observer position.  The  homogeneous atmosphere has by definition  
a constant air density $\rho$ equal to the one at observer position and its height is such that it 
would give the  same pressure as the one recorded at observer position. Note that we do not 
assume here that the atmosphere is homogeneous, we just use the reduced height that can be obtained
for any real atmosphere just from the  pressure and density at observer. Assuming furthermore 
ideal gas law for dry air we have:

 \begin{equation}
\label{eq:ell}
\ell(T)={P \over {\rho \ g}}={P_0 \over {\rho_0 \ g_0}}{T \over T_0},
\end{equation} 
where ${\rho_0=1.293\ \mbox{kg}\,\mbox{m}^{-3}}$ for ${T_0=273.15\,\mbox{K}}$, ${P_0=101325\,\mbox{Pa}}$ and normal gravity ${g_0=9.80665\,\mbox{m}\,\mbox{s}^{-2}}$. 
The  radius of curvature for Calern observatory (${\varphi=43\degr45\arcmin7\arcsec}$, ${h=1323\,\mbox{m}}$) was approximated by the minimum reference ellipsoid curvature at latitude $45\degr$ and sea level (\cite{Chollet1981PhD}, see Appendix~\ref{App:curvature}):
\begin{equation}
r_c(45\degr,0)=6367.512\ \mathrm{km}
\end{equation}
Ambient  air refractivity was 
deduced from the refractive index $n_0(\lambda)$ under standard conditions 
 and the partial pressure of 
water vapor $p$ by applying the formula recommended by the first resolution of the 13th General Assembly of the International Union of Geodesy and Geophysics (IUGG \citeyear{IUGG1963}; Baldini~\citeyear{Baldini1963}). After conversion to Pa (Pascal) 
as the pressure unit, the equation becomes:
\begin{equation}
\label{eq:alpha}
\alpha(T,P,f_h,\lambda)={T_0 \over T} \left\{(n_0(\lambda)-1) {P \over P_0} - 4.13\, 10^{-10}\ p(f_h,T)\right\}
\end{equation}
Refractivity under standard condition (sea level, ${T=T_0}$, ${P=P_0}$, $0\%$ humidity, $0.03\%$ of carbon dioxide) was taken from the work of \cite{Barrell&Sears1939}:
\begin{equation}
\label{eq:Barrel}
n_0(\lambda)-1=\left\{2876.04+{16.288 \over (10^6\lambda)^2}+{0.136 \over (10^6\lambda)^4}\right\}\ 10^{-7}.
\end{equation}
Partial pressure of water vapor for the current temperature and relative humidity 
was deduced from a fit of water vapor pressure data published by the Bureau Des Longitudes \citeyearpar{BDL1975} for temperatures between $-15\ {\degr}\mbox{C}$ and $+25\ {\degr}\mbox{C}$.
 The resulting equation, converted to Pa, is \citep{Chollet1981PhD}:
\begin{equation}
\label{eq:water}
p(f_h,T)=f_h\ 610.75\ e^{7.292\, 10^{-2}(T-T_0)-2.84\, 10^{-4}(T-T_0)^2}
\end{equation}
Finally, we note that local atmospheric pressure $P$ was measured from the height $H$ (in mm) 
of a mercurial barometer and its temperature $\theta$ (in $\degr\mbox{C}$). 
Taking into account corrections for local gravity (latitude and altitude) and for temperature 
 (through the volume thermal expansion of mercury and the coefficient of linear thermal expansion 
of the tube),  $P$ was obtained by\footnote{\cite{Chollet1981PhD} used erroneously $2.64\, 10^{-4}$ in this equation.} (see Appendix \ref{App:baro}):
\begin{equation}
\label{eq:baro}
P=H\left\{1-2.64\, 10^{-3}\cos(2\varphi)-1.96\, 10^{-7} h - 1.63\, 10^{-4}\ \theta\right\}
\end{equation}

\subsection{Error function formula}
In fact, in Eq.~(\ref{eq:Danjon}), only the first two terms which correspond to Laplace formula can be found without any hypothesis on the real atmosphere (only the reduced height $\ell$ and the refractivity at observer are needed). 
The term in $\tan^5$
comes from an additional assumption, namely the fact that air density follows an exponential decrease with height (actually with a well chosen variable which vary almost linearly with height).
This leads to the following equation (see \cite{Danjon1980, Fletcher1931}):
\begin{equation} 
\label{eq:erfc}
R=\alpha \left({{2-\alpha}\over {\sqrt{2\beta-\alpha}}}\right)\sin(z)\ \Psi\left({{\cos(z)}\over{\sqrt{2\beta-\alpha}}}\right)
\end{equation}
with :
\begin{equation}
\Psi(x)=e^{x^2}\int_x^{\infty}e^{-t^2}dt={{\sqrt{\pi}}\over{2}}e^{x^2}\left(1-\mathrm{erf}(x)\right)
\end{equation}
from which Eq.~(\ref{eq:Danjon}) was derived by keeping only the three first terms of its asymptotic expansion.

\subsection{Cassini}
By comparing the results with a full integration method, \cite{Young2004} shows the superiority of Cassini's formula over the series-expansion approach and advocates 
its use by astronomers.  \cite{Cassini1662} assumed an homogeneous atmosphere for which he obtained the exact formula:
\begin{equation}\label{eq:Cassini}
R=\mathrm{asin}\left({{n_{obs}\ r_c\sin(z)}\over{r_c+\ell}}\right)-\mathrm{asin}\left({{r_c\sin(z)}\over{r_c+\ell}}\right)
\end{equation}
The demonstration of this formula can also be found in  \cite{Young2004}. Again, it can be shown \citep{Ball1908} that expanding this formula also leads to the to first two 
terms of Eq.~(\ref{eq:Danjon}) i.e. to Laplace formula.

In Table~\ref{tab:absref} the three formulae discussed above are compared to the full numerical integration of Eq.~(\ref{eq:int}) with a standard atmospheric model for six observed zenith distances and average weather conditions at Calern observatory. This illustrates the use of the three approximate formulae which are fully determined by the evaluation of $\alpha$ and $\beta$ for the actual weather conditions at the observing station.
At $85^\circ$, the $\tan^5$ expansion diverge while Cassini's and the error function formulae remain closer to the result of the full numerical integration.  

\begin{table} 
\caption{Refraction in arcseconds as a function of the observed zenith distance for different approximations of the refraction integral. `tan$^5$' refers to Eq.~(\ref{eq:Danjon}), `Cassini' to Eq.~(\ref{eq:Cassini}), `Erf' to Eq.~(\ref{eq:erfc}) and `Full integration' to the full numerical integration of Eq.~(\ref{eq:int}) using a standard atmosphere model (see Section~\ref{sec:absval}). We took $\lambda=782.2$~nm, T=15\ {\degr}\mbox{C}, P=875~hPa, $f_h=50\%$. The corresponding air refractivity is $\alpha$=n$_{\mathrm{obs}}$-1=2.373 $10^{-4}$ and the reduced height is $\ell=8430$~m which corresponds to $\beta=0.00132$ at Calern observatory.  }
\label{tab:example}
\begin{tabular}{lcccc}
\hline
z   & $\tan^5$ & Cassini & Erf  & Full integration\\
\hline
$10^\circ$&   8.618'' &  8.618'' &  8.618'' &  8.618''\\
$30^\circ$&  28.207'' & 28.208'' & 28.208'' & 28.208''\\
$50^\circ$&  58.149'' & 58.149'' & 58.150'' & 58.150''\\
$70^\circ$& 133.097'' & 133.084'' & 133.104'' & 133.094''\\
$80^\circ$& 267.683'' & 267.054'' & 267.553'' & 267.411''\\
$85^\circ$& 512.185'' & 487.335'' & 495.780'' & 494.176''\\
\hline
\end{tabular}
\label{tab:absref}
\end{table}

%%%%%%%%%%%%%%%%%%%%%%%%%%%%%%%%%%%%%%%%%%%%%%%%%%%%%%%%%%%%%%%%%%%%%%%%%%%%%%
\section{On the observed shape of the Sun due to pure astronomical refraction}
\label{sec:shape}
%%%%%%%%%%%%%%%%%%%%%%%%%%%%%%%%%%%%%%%%%%%%%%%%%%%%%%%%%%%%%%%%%%%%%%%%%%%%%
\begin{figure}
   \centering
   \includegraphics[width=.4\textwidth]{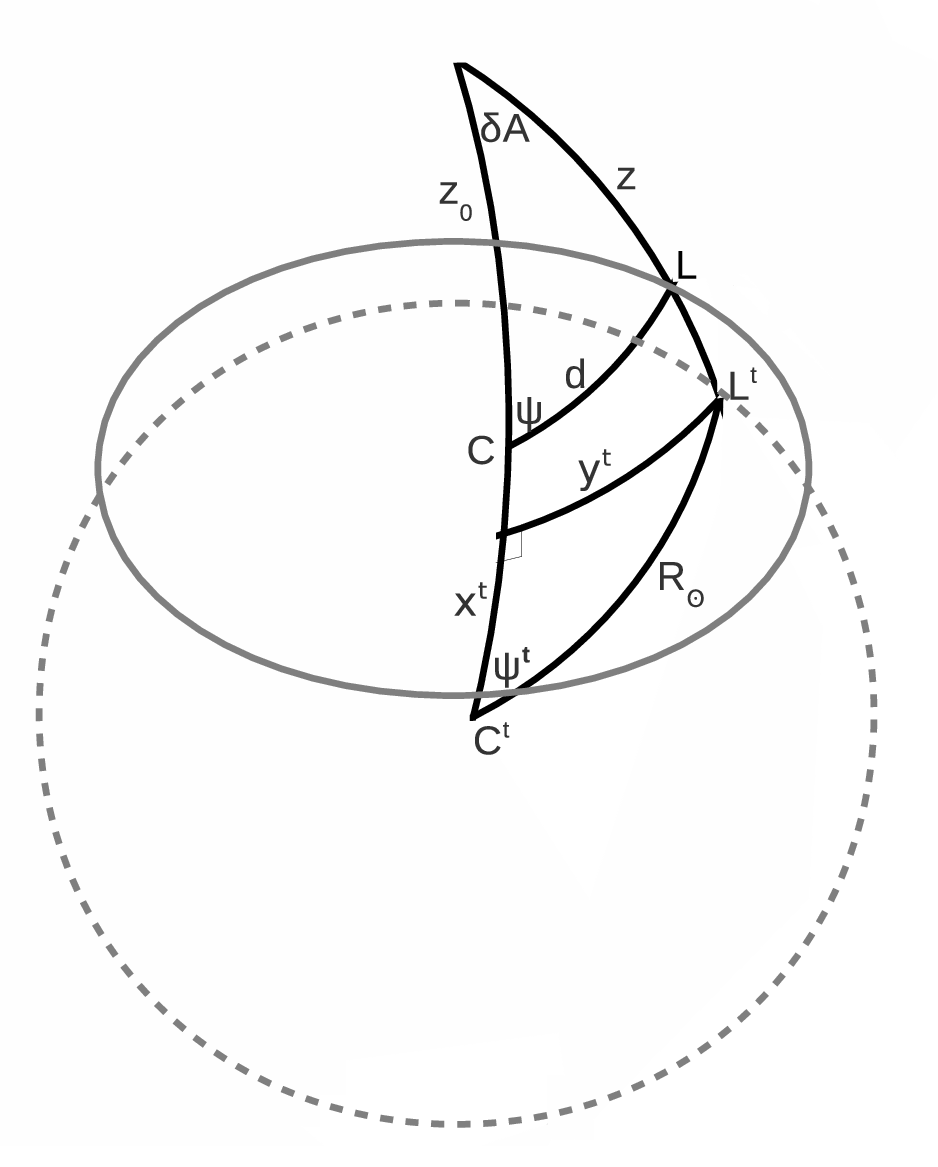}
      \caption{Geometry for the solar shape due to astronomical refraction.The dashed circle represents the true solar disk of centre C$^t$ and radius
			$R_{\sun}$ while the elliptical shape (full line) represents the observed Sun of centre C. The point at the top represents observer's zenith.
              }
         \label{Fig:Dessin}
   \end{figure}

In this section, we assume that the Sun is a perfect sphere of angular radius $R_{\sun}$ 
at 1 AU and that there is no other effect
affecting its observed shape than astronomical refraction defined by Eq.~(\ref{eq:gen_alt}).

 In the horizontal 
coordinate system (zenith distance-azimuth), we note  ($z_{\sun}^t$, $A_{\sun}$) the true position 
of the Sun centre ($C^t$) observed at zenith angle $z_{\sun}$;   ($z^{t}$, $A$) the true position 
of a point  ($L^t$) of the solar limb observed at zenith angle $z$; ${\delta z=z-z_{\sun}}$ and  ${\delta A=A-A_{\sun}}$.
Figure~\ref{Fig:Dessin} shows all the angles involved.
Each true limb point position can be defined by the angle $\psi^t\in[-\pi,\pi[$ between the direction $\overline{C^tL^t}$  
and the vertical circle. Similarly, each observed limb point can be located by the angle $\psi\in[-\pi,\pi[$ 
between the observed direction $\overline{CL}$  and the vertical circle.
However, because the figure is symmetric with respect to the vertical circle, we 
consider only the interval $[0,\pi]$ for $\psi$ and $\psi^t$ in the following. For observation with an Alt-Az mount this 
 would correspond directly to the angle with  one of the CCD axis. 
For an equatorial mount, one CCD axis is aligned with the hour circle passing through the celestial poles and the Sun and 
therefore the vertical circle can be materialized on the solar image by computing first the parallactic angle 
between these two circles.

If ${d(\psi)=\bar{d}(\psi^t)}$ is the angular distance between the observed position of the Sun centre
and the observed limb points, we define by:
\begin{equation}\label{eq:ddef}
<d>={1 \over {\pi}}\int_0^{\pi}d(\psi)\mathrm{d}\psi={1 \over {\pi}}\int_0^{\pi}\bar{d}(\psi^t)\mathrm{d}\psi^t
\end{equation}
the geometric  mean radius of the observed Sun. The horizontal and vertical angular extent of the observed Sun are noted 
$D_{h}$ and $D_{v}$ respectively and, the flattening is given by:
\begin{equation}\label{eq:flatdef}
f={{D_{h} - D_{v}} \over D_{h}}
\end{equation}
Following \cite{Mignard2010}, we define the magnification $\Gamma$ as the ratio between the vertical size of the image ($\delta z$)
of a small object to its true size ($\delta z^t$), i.e. in differential form $\Gamma\equiv dz/dz^t$. From Eqs. (\ref{eq:gen}) and (\ref{eq:gen_alt}), we have respectively:
\begin{equation}\label{eq:gamma}
\Gamma=\left({1+{{dR}\over{dz}}}\right)^{-1} \ \ \ \mbox{and}\ \ \ \ \Gamma=1-{{d\bar{R}}\over{dz^t}}
\end{equation}
The distorsion $\Delta$ is then defined as the rate of change of the magnification i.e. $\Delta\equiv {d\Gamma}/{dz}$. By derivating the two relations Eq.~(\ref{eq:gamma}) we obtain respectively:
\begin{equation}\label{eq:delta}
\Delta=-\Gamma^2{{d^2R}\over{dz^2}} \ \ \ \mbox{and}\ \ \ \ \Delta=-{1\over\Gamma}{{d^2\bar{R}}\over{d{z^t}^2}}
\end{equation}

%%%%%%%%%%%%%%%%%%%%%%%%%%%%%%%%%%%%%%%%%%%%%%%%%%%%%%%
\subsection{Approximate formulae for all zenith angles}
%%%%%%%%%%%%%%%%%%%%%%%%%%%%%%%%%%%%%%%%%%%%%%%%%%%%%%%%%
Any limb point true position can be located by its projections on the vertical circle passing through the true Sun centre, 
and on the great circle perpendicular to this vertical circle passing through the limb point (see Fig.~\ref{Fig:Dessin}). 
Because all the angles involved are small, we can write:
\begin{eqnarray}
x^t&=&R_{\sun}\cos(\psi^t)\\
y^t&=&R_{\sun}\sin(\psi^t)
\end{eqnarray}
and:
\begin{equation} \label{eq:circle}
{x^t}^2+{y^t}^2=R_{\sun}^2
\end{equation}
By looking at the expression of the observed values $x$ and $y$ of these projections,
one can obtain an approximate formula for the observed shape of the Sun.

The projection  $x^t$ on the vertical circle can  be approximated by keeping the two first terms of a Taylor expansion of the refraction:
\begin{equation}\label{eq:xt}
x^t\simeq z^t-z_{\sun}^t=\delta z+R(z)-R(z_{\sun})\simeq\delta z\left(1+ {{dR}\over{dz}}\right)+ {{(\delta z)^2}\over{2}}{{d^2R}\over{dz^2}}
\end{equation}
The observed projection $y$ is linked to $z$ and $\delta A$ both by the cosine and sine rules:
\begin{eqnarray}
\cos(y^t)&\simeq&\cos^2(z^t)+\sin^2(z^t)\cos(\delta A) \label{eq:cos}\\
\sin(y^t)&=&\sin(\delta A) \sin(z^t) \label{eq:sin}
\end{eqnarray}
Differentiating Eq.~(\ref{eq:cos}) and using Eq.~(\ref{eq:sin}) with ${\sin(y^t)\simeq y^t}$, ${\sin(\delta A)\simeq \delta A}$ and ${dz^t=-\bar{R}(z^t)}$ leads to:
\begin{equation}
dy^t={{-y^t\ \bar{R}(z^t)}\over{\tan(z^t)}}
\end{equation}
The observed distance $y$ is then obtained by:
\begin{equation} \label{eq:yt}
y\simeq \delta A \sin(z)=y^t+dy^t=y^t\left({ 1-{{\bar{R}(z^t)}\over{\tan(z^t)}}}\right)
\end{equation}

Finally, by reporting Eqs.~(\ref{eq:xt}) and (\ref{eq:yt}) in Eq.~(\ref{eq:circle}) and using Eqs.~(\ref{eq:gamma}) and (\ref{eq:delta}), we obtain:
\begin{equation}\label{eq:shape}
\left[{{\delta z}\over{\Gamma}}-{\Delta\over 2}\left({{\delta z} \over \Gamma}\right)^2\right]^2+\left[{{\delta A \sin z}\over{1-{{\displaystyle\bar{R}(z^t)}\over{\displaystyle\tan(z^t)}}}}\right]^2=R_{\sun}^2
\end{equation}
where the magnification and distortion are taken at $z_{\sun}$.
From this we can deduce the position of the two vertical limb points and the observed vertical extent of the image.
 For ${\Delta \ll R_{\sun}}$ and ${\delta A=0}$, we find:
\begin{eqnarray}
d(\pi)&\simeq&\Gamma\, R_{\sun} \left(1+{{\Delta R_{\sun}}\over 2}\right) \label{eq:dpi}\\
d(0)&\simeq&\Gamma\, R_{\sun} \left(1-{{\Delta R_{\sun}}\over 2}\right)\label{eq:d0}
\end{eqnarray}
and thus:
\begin{equation}\label{eq:Dv}
D_{v}=d(0)+d(\pi)\simeq 2\,\Gamma\, R_{\sun}
\end{equation}
In the horizontal direction we obtain from Eq.~(\ref{eq:shape}) with ${\delta z =0}$:
\begin{equation}\label{eq:approxdmax}
D_{h}=2d(\pi/2)\simeq  2R_{\sun} \left(1-{{\bar{R}\left(z_{\sun}^t\right)}\over{\tan\left(z_{\sun}^t\right)}}\right)
\end{equation}

%%%%%%%%%%%%%%%%%%%%%%%%%%%%%%%%%%%%%%%%%%%%%%%%%%%%%%%%%%%%%%%%%%%%%%%%%%%%
\subsection{Approximate formulae for small zenith angles - elliptic shape}
%%%%%%%%%%%%%%%%%%%%%%%%%%%%%%%%%%%%%%%%%%%%%%%%%%%%%%%%%%%%%%%%%%%%%%%%%%%%
Keeping only the first term in Eq.~(\ref{eq:Danjon}) is equivalent to neglecting
 Earth curvature.
We obtain the following approximation valid close to the zenith only :
\begin{equation}
R(z)=k\tan(z) \ \ \mathrm{with:} \ \ k=\alpha(1\!-\!\beta)
\end{equation}
For the conditions of Table~\ref{tab:absref} we have  $k\simeq49''$ (from Eqs.~(\ref{eq:alpha1}) and (\ref{eq:beta})). For sea level pressure $P_0$ (other parameters beeing unchanged) we would obtain  $k\simeq57''$.
For this flat-Earth approximation we can also write:
\begin{equation}\label{eq:kkp}
\bar{R}(z^t)\simeq k'\tan(z^t) \ \ \mathrm{with:} \ \ k'= k\big(1-k\ \mathrm{sec}^2(z^t)\big)
\end{equation}
In that case and if we neglect the distortion, Eq.~(\ref{eq:shape})
is reduced to the equation of a simple ellipse  (see also e.g.  Dionis Du S\'ejour~\citeyear{DuSejour1786}; Ball~\citeyear{Ball1908}): 
\begin{equation}
{{x^2}\over{\left(1-k'\ \mathrm{sec}^2\left(z_{\sun}^t\right)\right)^2}}+{{y^2}\over{(1-k')^2}}=R_{\sun}^2
\end{equation}
where ${x=\delta z}$ and ${y=\sin(z)\,\delta A}$ can be assimilated to Cartesian coordinates on 
two perpendicular axes on the image. 
The major axis of the observed ellipse is thus given by:
\begin{equation}\label{eq:dmax}
{D_{h}\over 2} =R_{\sun}(1-k')
\end{equation}
while the observed minor axis is:
 \begin{equation}\label{eq:dmin}
{D_{v}\over 2} =R_{\sun}\left(1-k'\ \mathrm{sec}^2\left(z_{\sun}^t\right)\right)
\end{equation}
We note from these equations that the Sun is shrunk in all directions. The  observed horizontal diameter 
is smaller than the true diameter but remains the same for all zenith angles (c.f. Fig.\ref{Fig:contract}) 
while the  observed vertical diameter
decreases with increasing zenith distance. The combination of these two effects  leads to the apparent 
flattening of the setting Sun (but keeping in mind that this approximate formula is not valid close to the horizon).
From Eqs.~(\ref{eq:flatdef}), (\ref{eq:dmax}) and  (\ref{eq:dmin}), the flattening for small zenith angles is:
\begin{equation}
f\simeq k \tan^2(z^t_{\sun}).
\end{equation}
while, near the horizon, Eq.~(\ref{eq:shape}) implies that the flattening is simply given by the vertical magnification
taken at the the Sun's centre.
For small zenith angles, the observed elliptic  shape can be written as:
\begin{equation}
d(\psi)={{D_{v}} \over { 2\sqrt{1-(2f-f^2)\sin^2(\psi)}}}
\end{equation}
which can be approximated by:
\begin{equation}\label{eq:dist_app}
d(\psi)\simeq R_{\sun}\left(1-k'\left(1+\cos^2(\psi)\tan^2(z^t_{\sun})\right)\right),
\end{equation}
and the mean radius is obtained by:
\begin{equation}\label{eq:meanell}
<d>\simeq R_{\sun} \left(1-k'-{k'\over 2}\tan^2(z^t_{\sun})\right)={{D_{v}+D_{h}}\over{4}}
\end{equation}
The validity of this approximation as a function of the zenith distance will be discussed in Section~\ref{sec:res} and checked against observations in Section~\ref{sec:data}.

%%%%%%%%%%%%%%%%%%%%%%%%%%%%%%%%%%%%%%%%%%%%%%%
\subsection{Exact formulae for all zenith angles}
%%%%%%%%%%%%%%%%%%%%%%%%%%%%%%%%%%%%%%%%%%%%%%%%
The classical approximate formulae above are useful for understanding the shape of the 
observed Sun in terms of magnification and distortion induced by refraction.
Equation (\ref{eq:shape}) shows that the general shape is a distorted ellipse with more flattening in the lower part 
than in the upper's. 
However, for a given refraction law, the shape of the observed Sun can also easily be obtained, in the general case, without any approximation.
In the following, we obtain first the solution of the forward problem: for  given true Sun radius $R_{\sun}$ and true 
zenith distance $z_{\sun}^t$, we obtain the shape of the observed Sun for any given refraction model. 
Then, we give the solution of the inverse problem: from the observed solar shape,  the knowledge of $z_{\sun}^t$
(from ephemeris)  and assuming a refraction model, we deduce the true angular solar radius.

\subsubsection{Forward problem}
Here we assume that the true zenith distance of the Sun centre 
$z_{\sun}^t$ and its true angular radius $R_{\sun}$ are known. For any refraction model $\bar{R}(z^t)$,
 and true angle $\psi^t$, we deduce the  observed angle $\psi$ and angular distance $d(\psi)$.
Applying the cosine and sine formulae respectively, we have :

\begin{equation}\label{eq:limb}
\left\{
\begin{array}{ l}
\!\!\!z^t=\mathrm{acos}\left[\cos\left(z_
{\sun}^t\right)\cos\left(R_{\sun}\right)+\sin\left(z_{\sun}^t\right)\sin\left(R_{\sun}\right)\cos(\psi^t)\right]\\
\\
\!\!\!\delta A=\mathrm{asin}\left({\displaystyle\sin(R_{\sun})\sin(\psi^t)}\over{\displaystyle\sin(z^t)}\right)
\end{array}
\right.
\end{equation}  
From Eq.~(\ref{eq:gen_alt}), we can get the observed zenith distances:
\begin{equation}\label{eq:limb2}
z=z^t-\bar{R}\left(z^t\right) \  \  \ \mathrm{and} \ \ \ z_{\sun}=z_{\sun}^t-\bar{R}\left(z_{\sun}^t\right)
\end{equation}  
and finally angular distances $\bar{d}(\psi^t)$ between the observed Sun centre and the observed positions of each limb point are obtained by application of the cosine rule:
\begin{equation}\label{eq:dist}
\bar{d}(\psi^t)\!=\!d(\psi)\!=\!\mathrm{acos}\big(\cos(z)\cos(z_{\sun})+\sin(z)\sin(z_{\sun})\cos(\delta A)\big)
\end{equation}
where the observed angle $\psi$ can be deduced from the true one by applying the sine rule:
\begin{equation}
\psi=\mathrm{asin}\left({{\sin(\delta A)\sin(z)}\over{\sin(\bar{d}(\psi^t))}}\right)=\mathrm{asin}\left({{\sin(z)}\over{\sin(z^t)}}{{\sin(R_{\sun})}\over{\sin(\bar{d}(\psi^t))}}\sin(\psi^t)\right)
\end{equation}

The smallest observed diameter of the Sun is obtained on the vertical direction:
\begin{equation}
D_{v}={d}(0)+{d}(\pi)=2R_{\sun}-\left(\bar{R}\left(z_{\sun}^t+R_{\sun}\right)-\bar{R}\left(z_{\sun}^t-R_{\sun}\right)  \right)
\end{equation}
and the largest angular extent, observed in the direction parallel to the astronomical horizon is obtained by:
\begin{equation}
D_{h}=2{d}(\pi/2)
\end{equation}
We note that  Eqs.~(\ref{eq:limb}) and (\ref{eq:limb2}) lead back to the approximation Eq.~(\ref{eq:approxdmax}) 
for the largest observed angular extent. This is however more easily obtained 
using the sine rule rather than Eq.~(\ref{eq:dist}). With ${\sin(d(\pi/2))\simeq d(\pi/2)}$, ${\sin(R_{\sun})\simeq R_{\sun}}$ and ${\cos(R_{\sun})\simeq 1}$, we obtain:
\begin{equation}\label{eq:sinerule}
{d}(\pi/2)\simeq\sin(z)\, \sin(\delta A)=\sin\left(z_{\sun}^t-\bar{R}\left(z_{\sun}^t\right)\right){{R_{\sun}}\over{\sin(z_{\sun}^t)}}.
\end{equation}
which, with a first order expansion of the sine function around $z_{\sun}^t$, leads to  Eq.~(\ref{eq:approxdmax}). 

%%%%%%%%%%%%%%%%%%%%%%%%%%%%%%%%%%%%%%%%%%%%%%%%%%
\subsubsection{Inverse problem}\label{Sec:inverse}
%%%%%%%%%%%%%%%%%%%%%%%%%%%%%%%%%%%%%%%%%%%%%%%%%%
Here we give the solution of the inverse problem: given a refraction model  ($R(z)$, $\bar{R}(z^t)$), knowing $z_{\sun}^t$ 
from ephemeris and the observed angular distance $d(\psi)$ between the observed Sun centre and a limb point at an 
observed angle  $\psi$ with the vertical circle, we deduce the true angular radius $R_{\sun}$.
For $\psi \ne 0$ and $\psi\ne\pi$, one can compute successively:
\begin{equation}\label{eq:inverse}
\left\{
\begin{array}{r  c l}
z_{\sun}&=&z_{\sun}^t-\bar{R}\left(z_{\sun}^t\right)\\
\\
\delta A&=&\mathrm{atan}\left[{{\displaystyle\sin(\psi)}\over{\displaystyle\sin(z_{\sun})\mathrm{cot}(d(\psi))-\cos(z_{\sun})\cos(\psi)}}\right] \\
\\
z&=&\mathrm{asin}\left[{{\displaystyle\sin(\psi)\sin(d(\psi))}\over{\displaystyle\sin(\delta A)}}\right]\\
\\
z^t&=&z+R(z)\\
\\
 \psi^t&=&\mathrm{atan}\left[{{\displaystyle\sin(\delta A)}\over{\displaystyle \sin\left(z_{\sun}^t\right)\mathrm{cot}(z^t)} - \displaystyle \cos\left(z_{\sun}^t\right)\cos(\delta A)}\right]\\
\\
R_{\sun}&=&\mathrm{asin}\left[{{\displaystyle\sin(\delta A)\sin(z^t)}\over{\displaystyle\sin(\psi^t)}}\right]\\
\end{array}
\right.
\end{equation}
For $\psi=0$ or $\psi=\pi$, we have: $\delta A=0$; $z=z_{\sun}\mp d(\psi)$; $\psi^t=\psi$ and $R_{\sun}=\pm\left(z_{\sun}^t-z^t\right)$.

%%%%%%%%%%%%%%%%%%%%%
\section{Results}\label{sec:res}
%%%%%%%%%%%%%%%%%%%%%%
\subsection{On the absolute value of refraction}\label{sec:absval}

We first look at the absolute value of refraction and compare the various 
approximate formulae of Section~\ref{sec:approx} to the full numerical 
integration of the refraction integral using a standard atmosphere \citep{Sinclair1982}.
This atmosphere is
assumed to be spherically symmetric, in hydrostatic equilibrium and made 
of a mixture of dry air and water vapor that follows the perfect gas law. 
It is made of two layers: the troposphere with a constant temperature gradient 
which extends from  the ground up to the tropopause at 11~km,  
and an upper isothermal stratosphere.
 Like in the US Standard Atmosphere \citep{US76}, the temperature 
and pressure at the surface are 
288.15~K and 101325~Pa and the constant tropospheric lapse rate is 
6.5~K~km$^{-1}$. In the troposphere, the relative humidity $f_h$ is assumed 
constant and equal to its value at the observer. 
The partial pressure of water vapor in a tropospheric layer at temperature $T$ is
then obtained by:
\begin{equation}\label{eq:vaporbis}
p(f_h,T)=f_h\left({T}\over{247.1}\right)^\delta 10^2 
\end{equation}
which, with ${\delta=18.36}$ \citep{Sinclair1982}, never depart by more than $0.5$~hPa from Eq.~(\ref{eq:water}) for temperature lower than $30\degr$. 
Dry air is assumed in the stratosphere.
Finally, Eq.~(\ref{eq:alpha}) and its derivatives with respect to $T$ and $P$ are used to find air 
refractivity  along the integral path.

The numerical integration was performed by using the method  of \cite{Auer&Standish2000} also recommended by the Astronomical Almanac 
\citep{Seidelmann1992}. The program used is based on the one 
published by \cite{Hohenkerk&Sinclair1985}) but adapted in order 
to use a dispersion equation based on the work 
of \cite{Peck&Reeder1972} in replacement 
of the less accurate equation of \cite{Barrell&Sears1939} (Eq.~(\ref{eq:Barrel})).
For the standard air defined by \cite{Ciddor1996} i.e.  ${T=15\ \degr\mbox{C}}$, ${P=P_0}$, $0\%$ humidity and $0.045\%$ of carbon dioxide, we take: 
\begin{equation} \label{eq:Peck}
 n_0(\lambda)-1=\left\{ { {0.05792105}\over{238.0185-\left(10^6\lambda\right)^{-2}} } + { {0.00167917}\over{57.362-\left(10^6\lambda\right)^{-2}} }  \right\}
\end{equation}	
This dispersion equation was also used by Ciddor \citeyearpar{Ciddor1996}
who derived a new set of equations for calculating the refractive index of air which was
 subsequently adopted by the International Association of Geodesy (IAG \citeyear{IAG1999})  
as a new standard.
 In the following, all computations have been made using 
$\lambda=782.2\ \mbox{nm}$
which is one of the wavelengths used by the PICARD-SOL project and that will be used in Section~\ref{sec:data}.

\begin{figure}
   \centering
 \includegraphics[width=0.31\textwidth,angle=90]{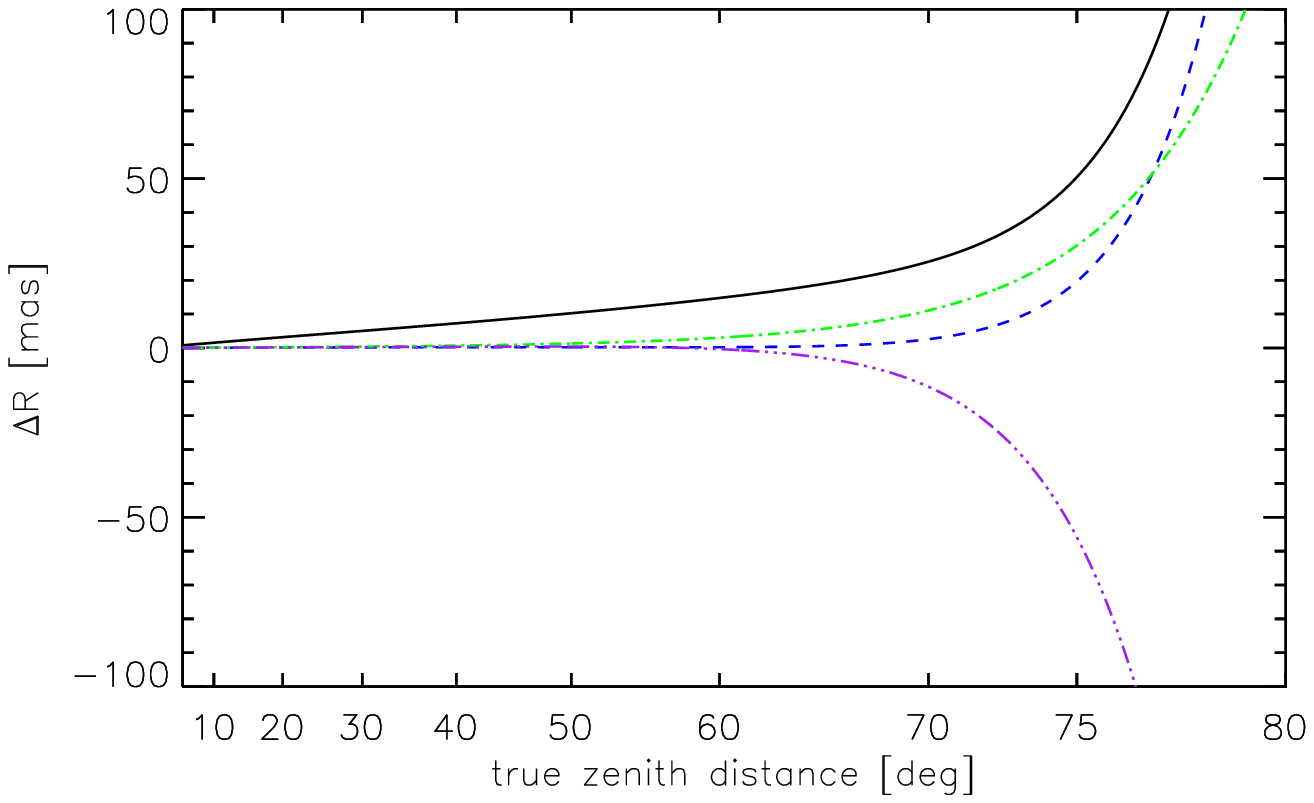}
      \caption{Absolute differences (in mas) between a reference model and the different approximate refraction formulae
 as a function of the true zenith distance. The reference model is obtained by full numerical integration of a
Standard Atmosphere \protect\citep{Sinclair1982} and Ciddor (\protect\citeyear{Ciddor1996}) equation for air refractivity. From top to bottom: $\tan^5$ expansion Eqs.~(\ref{eq:Danjon})-(\ref{eq:water}), full error function Eq.~(\ref{eq:erfc}), $\tan^5$ expansion Eq.~(\ref{eq:Danjon}), Cassini's formula Eq.~(\ref{eq:Cassini}). All approximate formulae but the top one use Ciddor (\protect\citeyear{Ciddor1996}) air refractivity.}
              
         \label{Fig:compabs}
   \end{figure}
   
\begin{figure}
   \centering
   \includegraphics[width=.31\textwidth,angle=90]{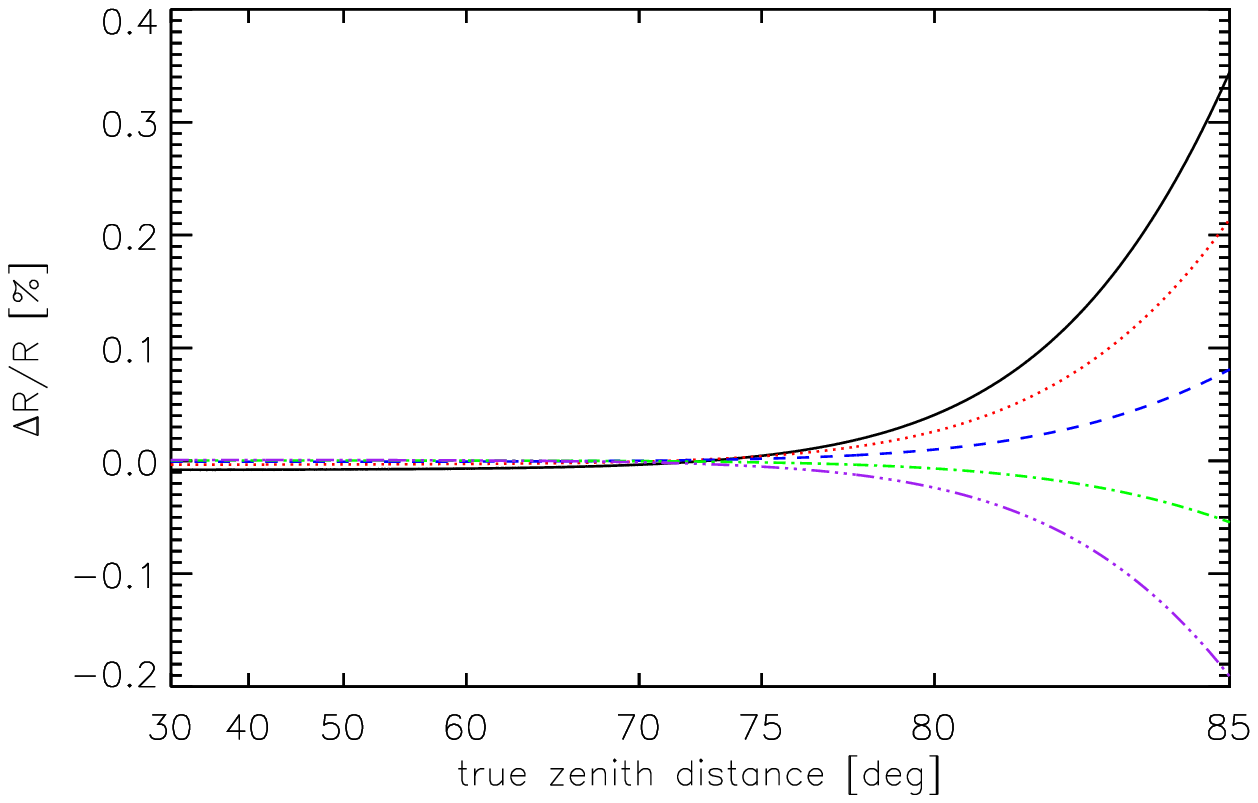}
      \caption{Relative error on refraction as a function of zenith distance for different tropospheric lapse rate. The reference model use the US Standard Atmosphere \protect\citep{US76} with a lapse rate of ${6.5\ \mbox{K\,km}^{-1}}$.
			The top curve correspond to an isothermal model and other atmosphere models have lapse rate of $2.5$, $5$, $7.5$ and ${10\ \mbox{K\,km}^{-1}}$  (from top to bottom at high zenith distance). 
			All models are computed using full numerical integration.}
              
         \label{Fig:comprel}
   \end{figure}

Figure~\ref{Fig:compabs} shows the absolute differences in 
milliarcseconds   (mas) 
between the approximate 
formulae and the exact integral evaluation for zenith distances up 
to $80\degr$. 
We immediately see that for  zenith distance lower than $75\degr$,
all the approximate formulae lead to less than 50~mas of absolute error. 
The full line corresponds to the $\tan^5$ formula Eq.~(\ref{eq:Danjon}) 
described in Section~\ref{sec:calern} while the dashed line corresponds 
to the same formula but using the new Ciddor~\citeyearpar{Ciddor1996} equations instead 
of Eqs.~(\ref{eq:alpha})-(\ref{eq:water}) for computing air refractivity. 
For zenith distances lower than  $80^\circ$,  the impact of using the old 
formula for refractivity never exceed 80~mas. The superiority of Ciddor 
equations to better fit observations and this for a wider range in wavelengths 
 is however clearly established.
The two
other lines correspond to the error function (dot-dash) and Cassini 
(triple dots-dash) formulae both using the Ciddor~\citeyearpar{Ciddor1996} equation 
for refractivity. These two last 
formulae were selected mainly because, unlike the series expansions in $\tan(z)$,
 they are finite at the horizon.
The full integration with standard atmosphere conditions 
leads to a refraction of about $1980\arcsec$ at the horizon. 
The error function and Cassini formulae lead respectively to $2088\arcsec$ 
and $1180\arcsec$ corresponding to relative errors of $5\%$
and $40\%$ respectively. This tends to favour the use of the error function 
formula over Cassini's one very close to the horizon. The hypothesis made 
to derive the error function formula are indeed more realistic than Cassini's 
hypothesis of an homogeneous atmosphere. 
It has however been shown that refraction
below $5\degr$ of the horizon is variable and  strongly depend
on the local lapse rate and properties of the boundary layer above or below
 the observer's eye (e.g. Young \citeyear{Young2004}). Within few degrees from the horizon, 
refraction may be influenced by thermal inversion boundary layers, ducting 
or other phenomena leading to extreme refraction. In this range, the local
lapse rate must be known and 
it is not expected that any formula using just the temperature and pressure 
at observer could give an accurate absolute refraction. 

It is however 
probably more interesting to look in the range between 
$60\degr$ and $80\degr$ of 
zenith distance, which is more important to astronomers willing 
to push in that range the limits of their astrometric measurements using
only temperature and pressure recorded at observer position.
We first note from Fig.~\ref{Fig:compabs} that, 
between $60\degr$ and $77\degr$, 
the $\tan^5$ expansion formula is actually giving slightly better absolute 
refraction values than the error function formula. If we now assume that
temperature and pressure at observer position are perfectly known, the only 
remaining important unknown in the atmospheric model is the tropospheric lapse 
rate. 
 We can however fix limits for a realistic lapse rate: it must lie between an isothermal model
and a lapse rate of ${10\ \mbox{K\,km}^{-1}}$ which would correspond to an adiabatic 
atmosphere \citep{Young2004}. Figure~\ref{Fig:comprel} shows  the relative 
error for such models with lapse rate ranging from 0 to ${10\ \mbox{K\,km}^{-1}}$ when 
they are compared to the standard model with a lapse rate of ${6.5\ \mbox{K\,km}^{-1}}$.
From this we can deduce that, no matter what is the real atmosphere, 
if the conditions at observer are known,  the relative error on refraction is lower than 
$0.01\%$ for zenith angles below $77\degr$ and lower than $0.4\%$ for zenith angles
between  $77\degr$ and  $85\degr$.

%%%%%%%%%%%%%%%%%%%%%%%%%%%%%%%%%%%%%%%%%%%%%%%%%
\subsection{On the mean solar radius correction}\label{Sec:mean}
%%%%%%%%%%%%%%%%%%%%%%%%%%%%%%%%%%%%%%%%%%%%%%%%%%

\begin{figure}
   \centering
\includegraphics[width=.31\textwidth,angle=90]{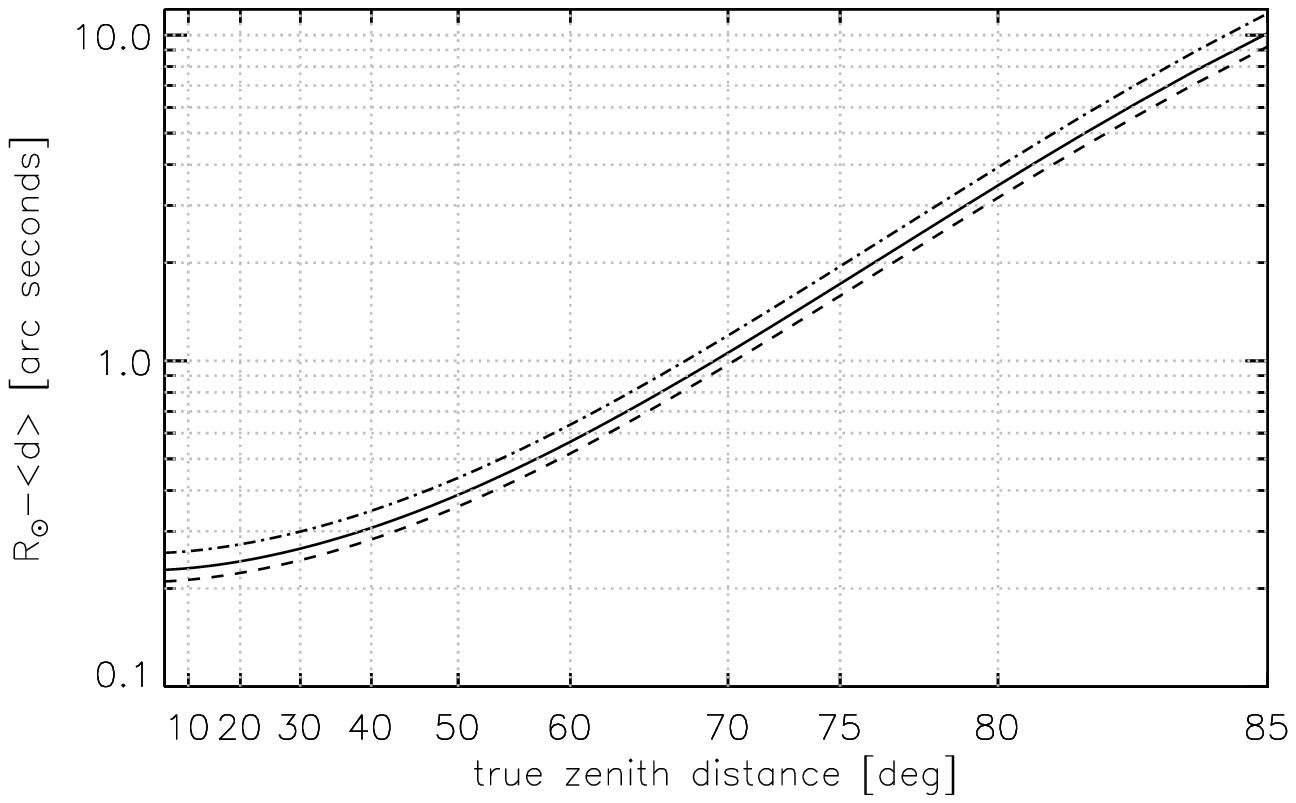}
      \caption{Difference between the true solar radius and the mean observed one as a function of the true zenith distance. The full line corresponds to average weather conditions at Calern  (${T=15\ {\degr}\mbox{C}}$,  ${P=875\ \mbox{hPa}}$). 
			The dot-dashed and dashed lines correspond respectively to ${T=-10\ {\degr}\mbox{C}}$,  ${P=900\ \mbox{hPa}}$ and  ${T=30\ {\degr}\mbox{C}}$,  ${P=850\ \mbox{hPa}}$.
			All calculations are made using the exact formulae Eqs.~(\ref{eq:ddef}) and (\ref{eq:dist}) for Calern station assuming $50\%$ humidity.}
              
         \label{Fig:refrac2diam}
   \end{figure}
\begin{figure}
   \centering
   \includegraphics[width=.31\textwidth,angle=90]{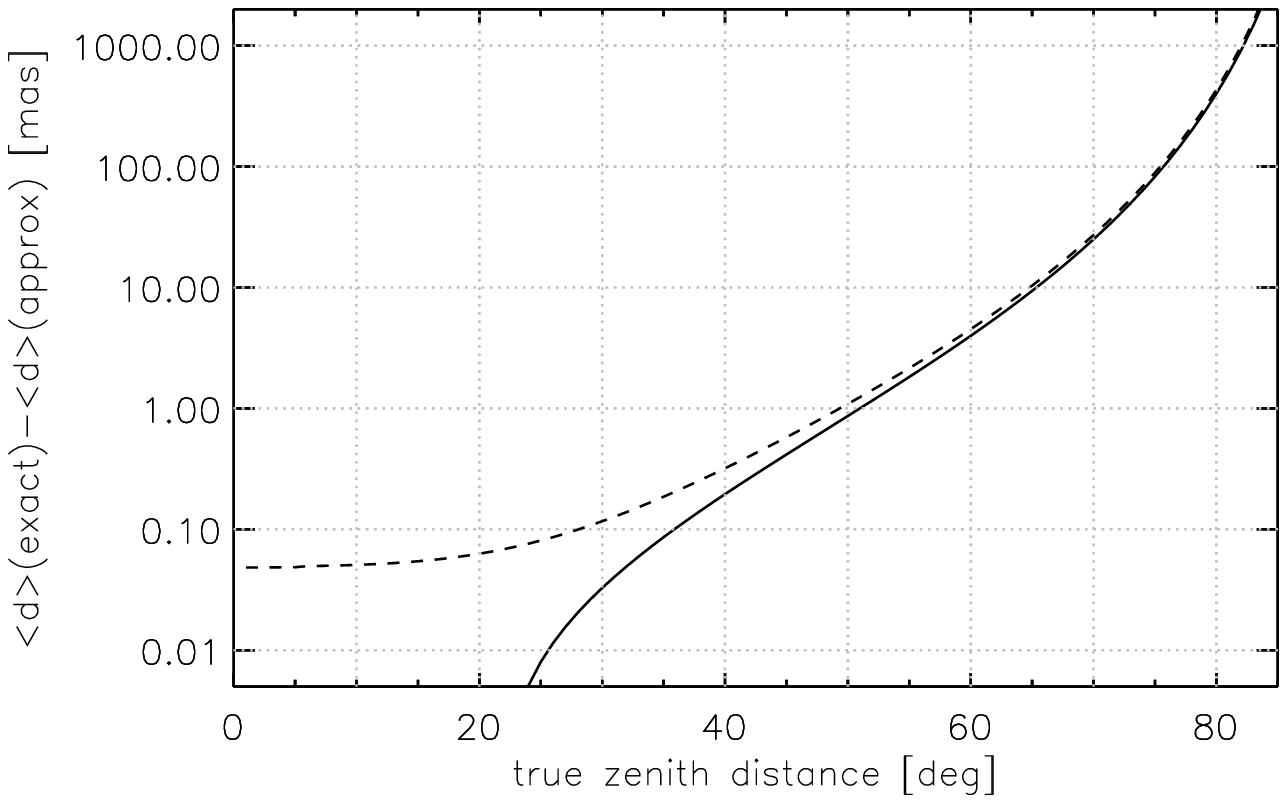}
      \caption{Difference between the correction due to refraction 
on the mean solar radius as calculated from integrating the exact formula Eq.~(\ref{eq:dist}) or using the approximate formula Eq.~(\ref{eq:meanell}). The dashed line is obained by replacing $k'$ by $k$ in  Eq.~(\ref{eq:meanell})}              
         \label{Fig:refrac2diam2}
   \end{figure}

Figure~\ref{Fig:refrac2diam} shows  the difference between the true radius of 
the Sun and the mean radius of the observed Sun as defined by Eq.~(\ref{eq:ddef}) as a function of the true zenith distance of the centre of the Sun. 
The exact formula Eq.~(\ref{eq:dist}) was used and we took standard conditions for Calern observatory (${T=15\ {\degr}\mbox{C}}$,  ${P=875\ \mbox{hPa}}$).
 The dashed and dot-dashed lines  are for ${T=-10\ {\degr}\mbox{C}}$,  ${P=900\ \mbox{hPa}}$ and  ${T=30\ {\degr}\mbox{C}}$,  ${P=850\ \mbox{hPa}}$
 respectively in order to illustrate the maximum amplitude of the effect at Calern station. 
The difference in the mean radius correction between the two extreme 
weather conditions range from 50~mas at the zenith up to 1850~mas 
at ${z^t=85\degr}$. It reaches 100~mas around ${z^t=55\degr}$ and 
200~mas around ${z^t=70\degr}$. This represents always less than $0.2\%$ of the correction.

Figure~\ref{Fig:refrac2diam2} shows the difference between the exact 
formula obtained by integrating Eq.~(\ref{eq:dist}) and the approximate formula Eq.~(\ref{eq:meanell}) corresponding to an elliptical shape.
The dashed line illustrates the result if $k'$ is approximated by $k$ (see Eq.~(\ref{eq:kkp})). In both cases the difference remains less than 20~mas for zenith distances lower than $70\degr$.
For larger zenith distances however, errors increase rapidly and the refraction function
 should be evaluated using full numerical integration.

%%%%%%%%%%%%%%%%%%%%%%%%%%%%%%%%%%%%%%%%%%%%%%%%%%%%%%%%%%%%%%%%%%%%
\subsection{On the angular dependence of solar radius correction} 
%%%%%%%%%%%%%%%%%%%%%%%%%%%%%%%%%%%%%%%%%%%%%%%%%%%%%%%%%%%%%%%%%%%%
\begin{figure}
   \centering
  \includegraphics[width=.31\textwidth,angle=90]{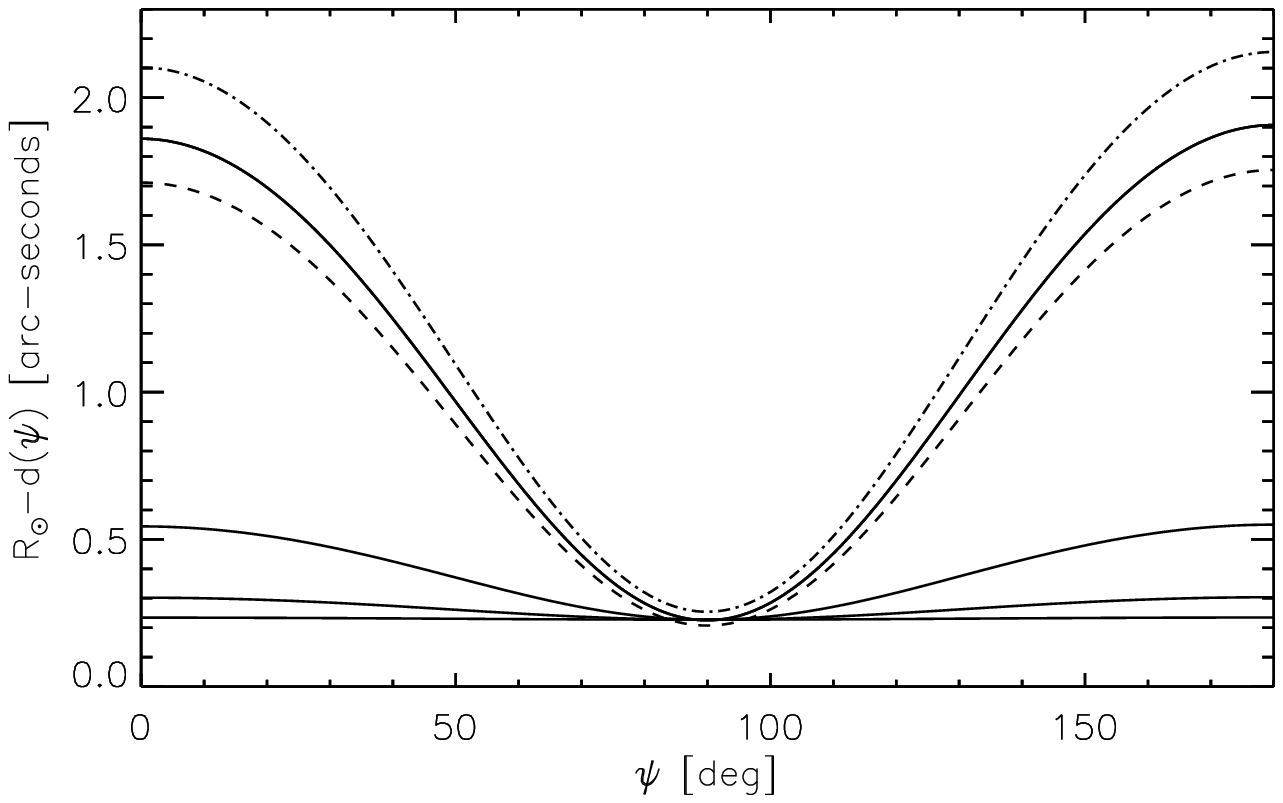}
      \caption{Difference between the true solar radius and the angular distances between the observed Sun centre and the observed positions of each limb points between the vertical
			(north for ${\psi=0\degr}$ and south for ${\psi=180\degr}$) and the horizon (${\psi=90\degr}$). The full lines are for ${z_{\sun}^t=70\degr}$, $50\degr$, $30\degr$ and $10\degr$ 
			respectively from top to bottom and are for average weather conditions at Calern. 
			The dashed and dot-dashed lines are for ${z_{\sun}^t=70\degr}$ and the same extreme weather conditions as in Fig.~\ref{Fig:refrac2diam}.}
         \label{Fig:refrac2diam3}
   \end{figure}
   \begin{figure}
   \centering
 \includegraphics[width=.31\textwidth,angle=90]{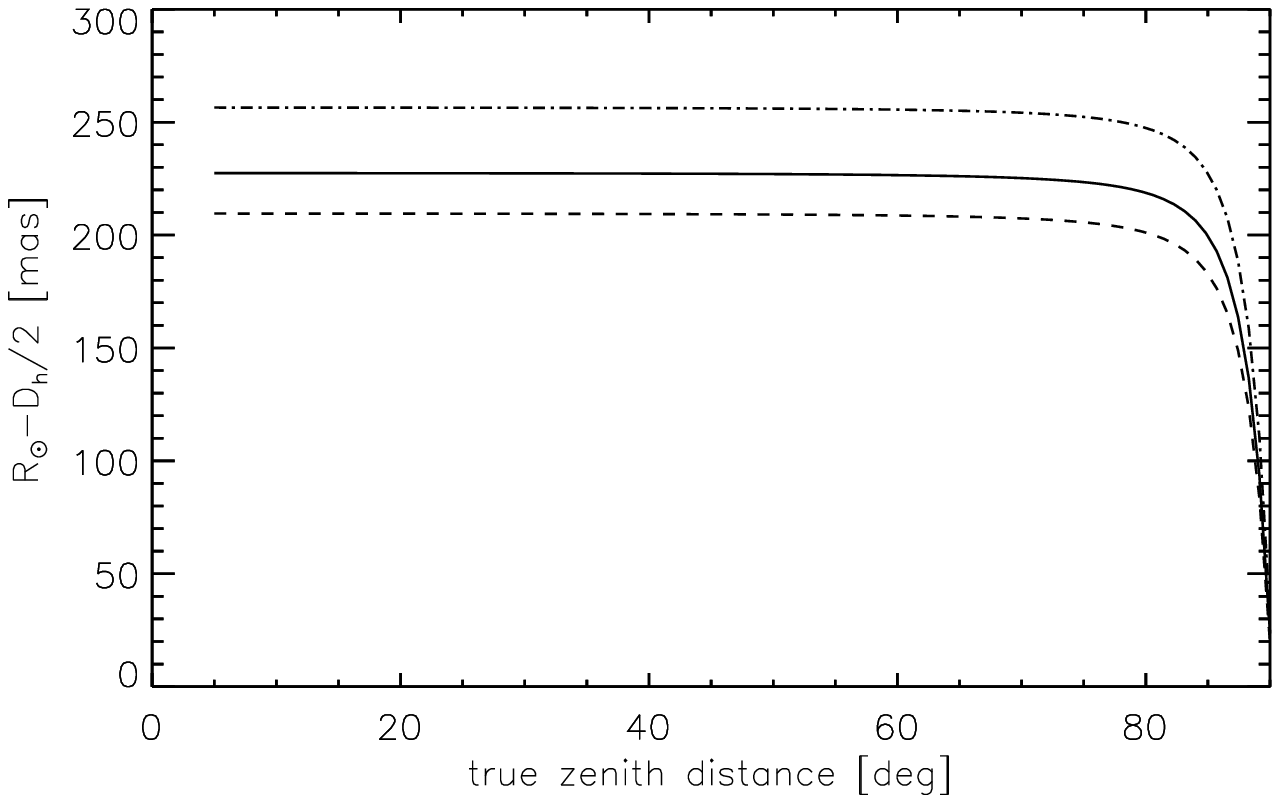}
      \caption{Contraction of the horizontal radius (${R_{\sun}-d(\pi/2)}$) as a function of the true zenith distance $z^t_{\sun}$. The full line is for average weather conditions at Calern. The dashed and dot-dashed lines are for  the same extreme weather 
			conditions as in Fig.~\ref{Fig:refrac2diam}.}         
         \label{Fig:contract}
   \end{figure}
For precise metrologic measurements of the Sun and in order to correct for other
effects (optical aberrations, turbulence, etc.) that are dependent on the position
 on the image, one may want to correct not the mean radius but each individual radius measured at all angles $\psi$. This can be done by following the 
procedure given in Section~\ref{Sec:inverse}. Figure~\ref{Fig:refrac2diam3} is obtained from Eq.~(\ref{eq:dist}) and  
illustrates the amplitude of the correction as a function of $\psi$ for different values of $z_{\sun}^t$, the true zenith distance of the Sun centre.
 We see that the horizontal diameter (${\psi=90\degr}$) is affected by refraction (by about ${2\times 0.23\arcsec=0.46\arcsec}$ for the chosen weather conditions) in agreement with Eq.~(\ref{eq:approxdmax}).
 The north and south vertical corrections (${\psi=0\degr}$ and $180\degr$ respectively) are also slightly
different in agreement with Eqs.~(\ref{eq:dpi})-(\ref{eq:d0}). Figure~\ref{Fig:contract} shows that the contraction of the horizontal radius lies between 210 and 260~mas depending on the actual weather conditions and remains constant
 for all zenith distances below $80\degr$ in agreement with Eq.~(\ref{eq:dmax}). It then decreases rapidly towards zero at the horizon  as expected from Eq.~(\ref{eq:approxdmax}). Physically the horizontal contraction results from the fact that meridians are not parallel lines (they cross at zenith). Near the horizon however they become parallel. 

%%%%%%%%%%%%%%%%%%%%%%%%%%%%%%%%%%%%%%%%%%%%%%%%%%%%%%%%%%%%%%%%	
\subsection{On uncertainties associated to radius corrections}
%%%%%%%%%%%%%%%%%%%%%%%%%%%%%%%%%%%%%%%%%%%%%%%%%%%%%%%%%%%%%%%%
\begin{figure}
   \centering
   \includegraphics[width=.31\textwidth,angle=90]{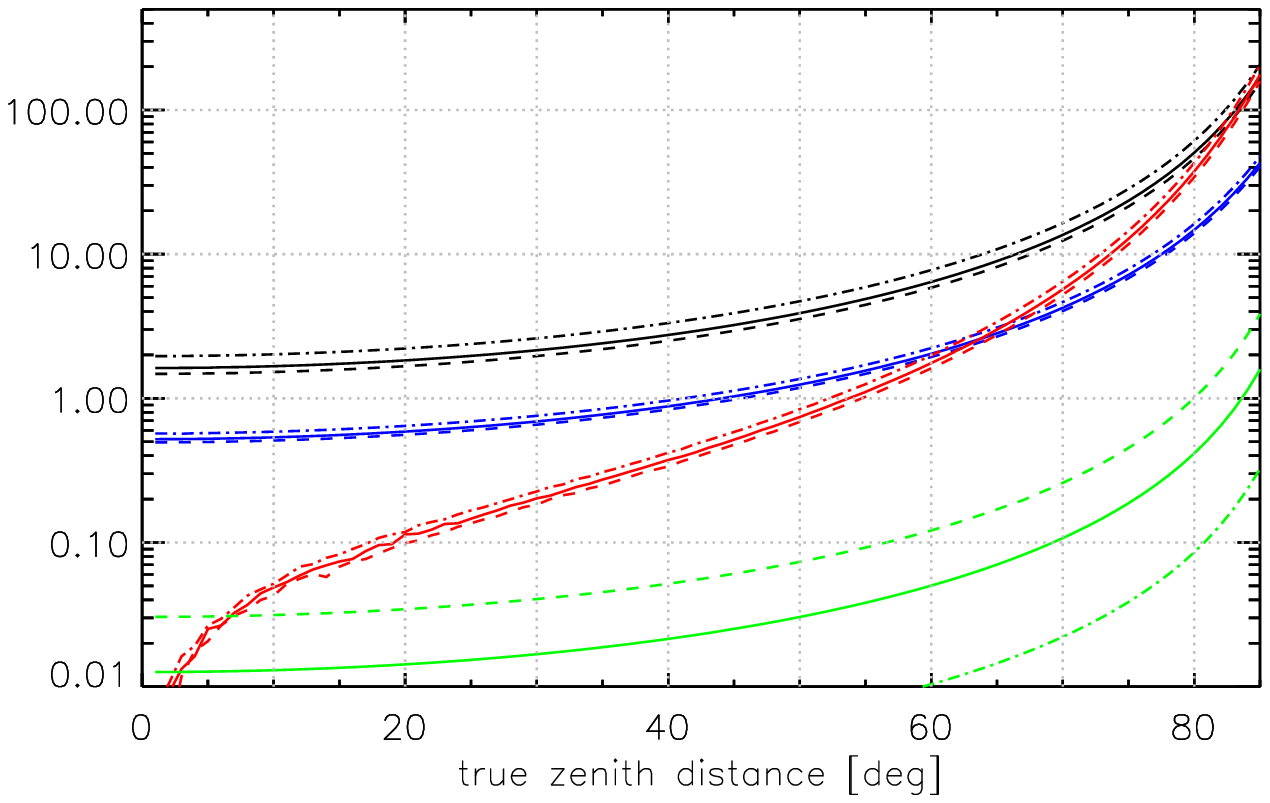}
      \caption{Partial derivatives of the vertical diameter correction ($\partial D_v/\partial X$) as a function of the true zenith distance.
			Partial derivatives in temperature, pressure, zenith distance and relative humidity are given 
			in $\mbox{mas\,K}^{-1}$, $\mbox{mas\,hPa}^{-1}$, $\mbox{mas\,arcmin}^{-1}$ and $\mbox{mas/\%}$ from top to bottom (at $40\degr$) respectively.
			The full, dashed and dot-dashed lines are for the same 
			weather conditions as on Fig.~\ref{Fig:refrac2diam}.}
              
         \label{Fig:refrac2diam5}
   \end{figure}
\begin{figure}
   \centering
  \includegraphics[width=.31\textwidth,angle=90]{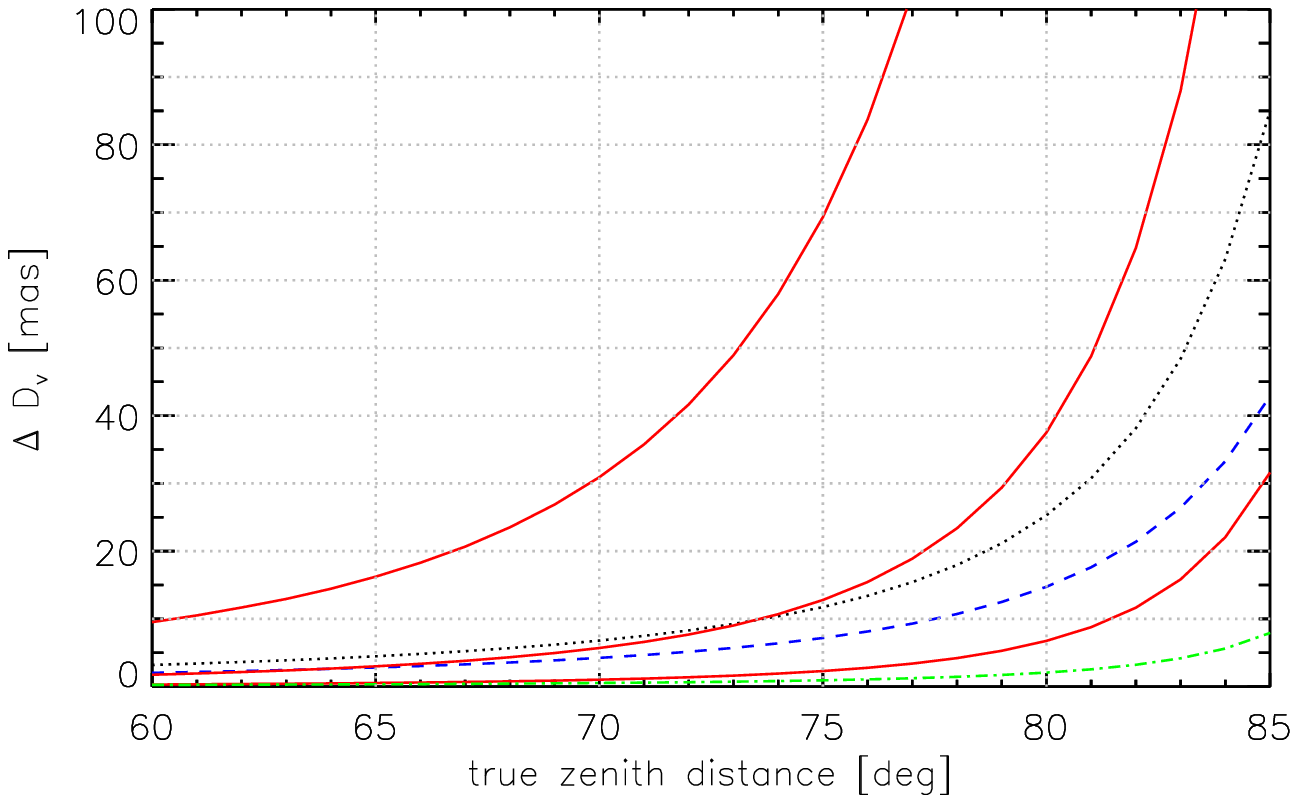}
      \caption{Uncertainties on the vertical diameter correction assuming ${\Delta T=0.5\ \mbox{K}}$ (dotted line), ${\Delta P=1\ \mbox{hPa}}$
			(dashed line), ${\Delta f_h=5\%}$ (dot-dashed line) and ${\Delta z_{\sun}^t=5.4\arcmin}$, $1.0\arcmin$ or $0.2\arcmin$  (full lines from top to bottom). The total error
			is obtained by summing the four contributions.}
              
         \label{Fig:refrac2diam6}
   \end{figure}

	We have shown that, apart from  weather conditions at observer's position, differences in atmospheric models and especially 
	different tropospheric lapses rate will not play any significant role at least up to $85\degr$ of zenith distance. The four 
	main contributions are therefore uncertainties in temperature, pressure, humidity and, for large zenith distance, uncertainties on the 
	true zenith distance itself.
	\begin{equation}\label{eq:deriv0}
	\Delta d(\psi)= \sqrt{\sum_{i=1}^4\left|{{\partial d(\psi)}\over{\partial X_i}}\right|^2\Delta X_i ^2}\ \ \ \ \ \ \ X=\left\{T,P,f_h,z_{\sun}^t\right\}
	\end{equation}
	It should be noted that we assume here observations made using filters with a narrow bandwidth around $\lambda$. For broadband filters, an additional term ${{\partial d(\psi)}/{\partial \lambda}}$ should be added
	by differentiating Eq.~(\ref{eq:Peck}).
	The largest uncertainty will be obtained for the vertical diameter (${D_v=d(0)+d(\pi)}$) which is the most affected by refraction. 
	Figure~\ref{Fig:refrac2diam5} shows the four partial derivatives contributing to $\Delta D_v$ between the two extreme weather conditions chosen above for Calern (see Section~\ref{Sec:mean}). 
	The partial derivatives shown have been obtained by numerically differentiating Eq.~(\ref{eq:dist}) but we have also checked that the analytical expressions that can be derived from the approximate 
	elliptical shape Eq.~(\ref{eq:dist_app}) are actually valid up to $80\degr$ of zenith distance.  Closer to the horizon the partial derivative over the zenith distance
	becomes significantly overestimated (c.f. Fig.~\ref{Fig:refrac2diamfinal}).
	From Eq.~(\ref{eq:dist_app}) and taking $k'\simeq k$, we obtain:
	\begin{equation}\label{eq:deriv1}
         \left\{
         \begin{array}{r c l} 
	\left|{{\displaystyle\partial d(\psi)}\over{\displaystyle\partial X_i}}\right|&=&\left|{{\displaystyle\partial k}\over{\displaystyle\partial X_i}}\right|\left(1+\cos^2(\psi)\tan^2\left(z^t_{\sun}\right)\right)R_{\sun} \ \ \ \ \ i=1..3\\
	\\
	\left|{{\displaystyle\partial d(\psi)}\over{\displaystyle\partial z^t_{\sun}}}\right|&=&2 k \cos^2(\psi){\rm sec}^2\left(z^t_{\sun}\right)\tan\left(z^t_{\sun}\right)R_{\sun}
	\end{array}
\right.
\end{equation}
	and from Eqs.~(\ref{eq:beta}), (\ref{eq:ell}), (\ref{eq:alpha}), (\ref{eq:kkp}), (\ref{eq:vaporbis}), we obtain: 
	\begin{equation}\label{eq:deriv2}
\left\{
         \begin{array}{r c l} 
	{{\displaystyle\partial k}\over{\displaystyle\partial T}}&=&-C_1{{P}\over{T^2}}(n_0(\lambda)-1)+C_3 T^{\delta-1}f_h\left(\delta C_2-{{\delta-1}\over{T}}\right)\\
\\
	{{\displaystyle\partial k}\over{\displaystyle\partial P}}&=&C_1(T^{-1}-C_2)(n_0(\lambda)-1) \\
\\
	{{\displaystyle\partial k}\over{\displaystyle\partial f_h}}&=&-C_3(T^{-1}-C_2)T^\delta\\
		\end{array}
\right.
\end{equation}
	where:
	\begin{equation}\label{eq:deriv3}
	C_1=T_0/P_0,\ \ C_2^{-1}=C_1r_c\rho_0g_0, \ \ C_3=4.13\,10^{-8}T_0(247.1)^{-\delta}
	\end{equation}
	For temperature, pressure and humidity, 
	we  assume uncertainties of ${\Delta T = 0.5\ \mbox{K}}$, ${\Delta P = 1\ \mbox{hPa}}$ and ${\Delta f_h = 5\%}$ which are typical for a standard weather station. 
	The precision on the true zenith distance relies on ephemeris calculations and a correct timing. At any given time ephemeris
	can give not only $z_{\sun}^t$ but also the instantaneous rate ${\mathrm{d}z_{\sun}^t}/{\mathrm{d}t}$ and, from the knowledge of  the image exposure time $\Delta t$,   one can deduce an uncertainty on $z_{\sun}^t$ by:
	\begin{equation}\label{eq:deriv4}
	\Delta z_{\sun}^t=\left|{{\mathrm{d}z_{\sun}^t}\over{\mathrm{d}t}}\right| \Delta t
	\end{equation}
	The maximum rate is about $650\arcsec\,\mbox{min}^{-1}$ at summer solstice. 
	Image exposures of 1~s, 5.5~s or 30~s would then correspond to a maximum uncertainty  $\Delta z_{\sun}^t$ of  $0.18\arcmin$,  
	 $1\arcmin$ or $5.4\arcmin$ respectively. These values are taken as illustrative examples. In the case of SODIM-II images , we use attenuation filter of about 20, the spectral filters are narrow (0.5-6.4~nm) and the exposure times range from 1.28~s to 8.90~s depending on the filter \citep{Meftah2014A&A}.
	Figure~\ref{Fig:refrac2diam6} shows
	the contribution of these uncertainties to the total uncertainty on vertical diameter correction for large zenith distances.
	We can see for instance that for $1\arcmin$ precision on the zenith distance (or 5.5~s exposure),
	the uncertainty coming from zenith distance can become, above $70\degr$,
		of the same importance as the combined uncertainties coming from temperature and pressure records. 
The total relative error
on the vertical diameter correction (${\Delta D_v/(2R_{\sun}-D_v)}$) remains however below $1\%$ up to ${z^t_{\sun} = 85\degr}$.

%%%%%%%%%%%%%%%%%%%%%%
\section{Application to real data}
\label{sec:data}
%%%%%%%%%%%%%%%%%%%%%%

The solar radius is very accurately determined for low zenith distances and the measurements obtained at high zenith distances corrected from the differential refraction effect should agree with this  value. Solar astrometry is therefore a very good way to test the validity of a differential refraction model.
For this,  we analysed a serie of 31711  near infrared solar images (782.2~nm and 1025.0~nm) recorded  at zenith distances ranging from $20\degr$  to $80\degr$. This includes 7978 images recorded at zenith distances above $60\degr$ which are normally excluded 
from our solar astrometry pipeline.
 These observations were performed at Calern Observatory using SODISM-II solar imager~\citep{Meftah2014A&A}. We selected these two wavelengths because the atmospheric turbulence effects, which also affect astrometric measurements, are much lower in the infrared than in the visible range.

For each image, after the usual CCD radiometric calibrations, we detect  the inflexion points of the center to limb profiles and thereby  extract 3600 values of the solar radius as a function of the heliographic angle. Because the SODISM-II mount is equatorial, we obtain  the observed angle $\psi$ for each inflexion point using the parallactic angle given by the ephemeris. Meteorological data (P, T, $f_h$, wind speed) are recorded simultaneously for each imge by a weather station located close to the instrument.

\begin{figure}
   \centering
\includegraphics[width=.5\textwidth]{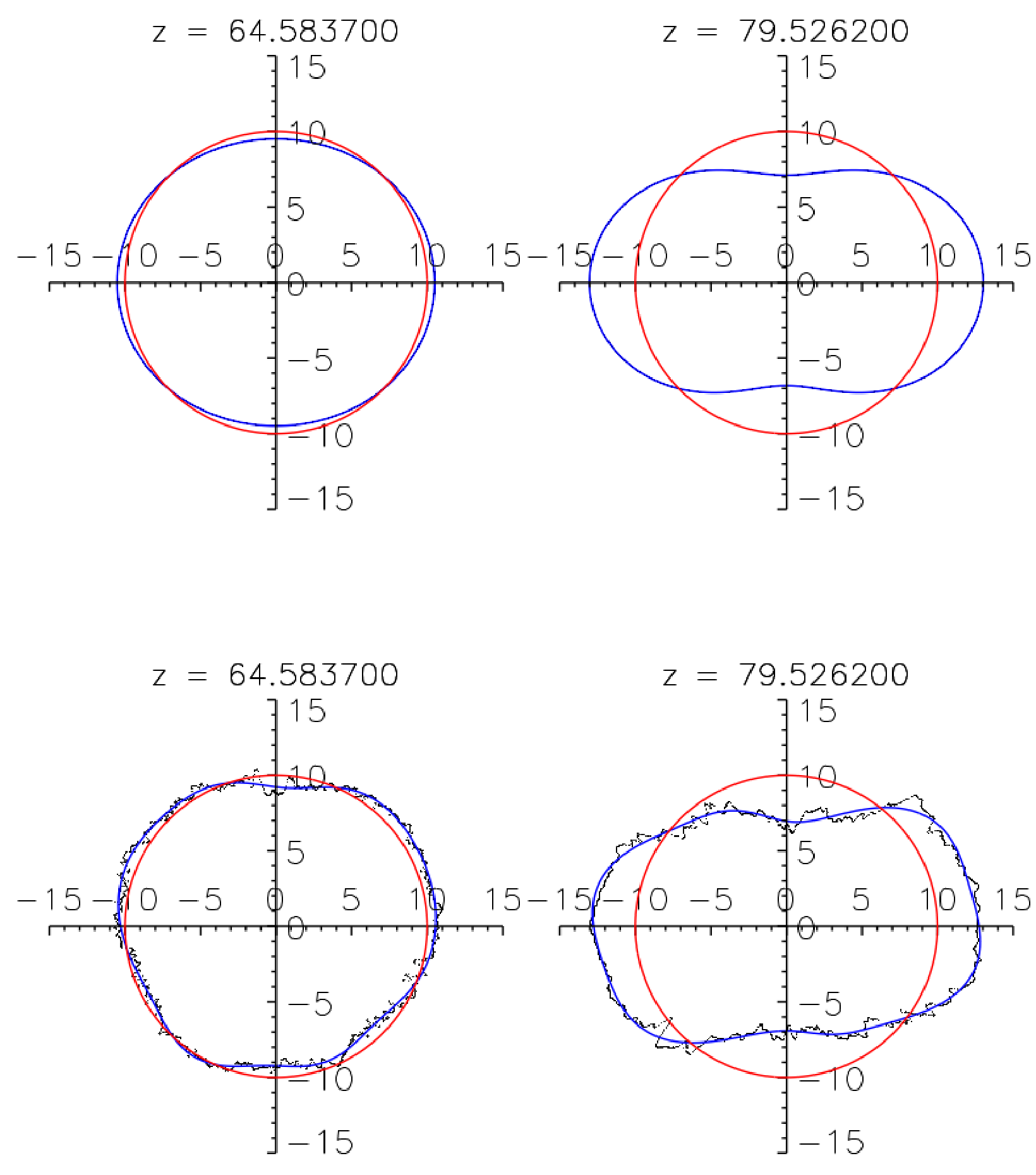}

    \caption{Observed solar shape (bottom) and the theoretical ones (top). These plots give two examples of solar shape observed at 782~nm and z=64.6\degr (left) or $z=79.5\degr$ (right) with the SODISM-II ground based instrument on november $25^{th}$ 2013. Black dots give the measured inflexion points of the observed limb darkening function. These are Fourier filtered (8 terms) to produce the blue lines which can be compared to the theoretical blue lines shown on the  top figures. The red circles give the mean of the observed refracted radii in each case for reference. In order to visualize the small differences from the mean radius we rescaled the mean radius to a value of 10 arcseconds while keeping the differences to the true scale i.e we plot   $d(\psi)-<d>+10$'' for the theoretical curves (top) and $R_\mathrm{obs} -<R_\mathrm{obs}>+10$''  for the observations (bottom). The scale on both axes give the distance from the mean radius in arc-seconds. The horizontal refraction being smaller than the vertical refraction, the observed horizontal diameter is larger than the mean observed one while the vertical diameter is smaller than the mean observed one.} 
              
         \label{Fig:refrac_sodism2}
   \end{figure}

\begin{figure}
   \centering
   \includegraphics[width=.475\textwidth]{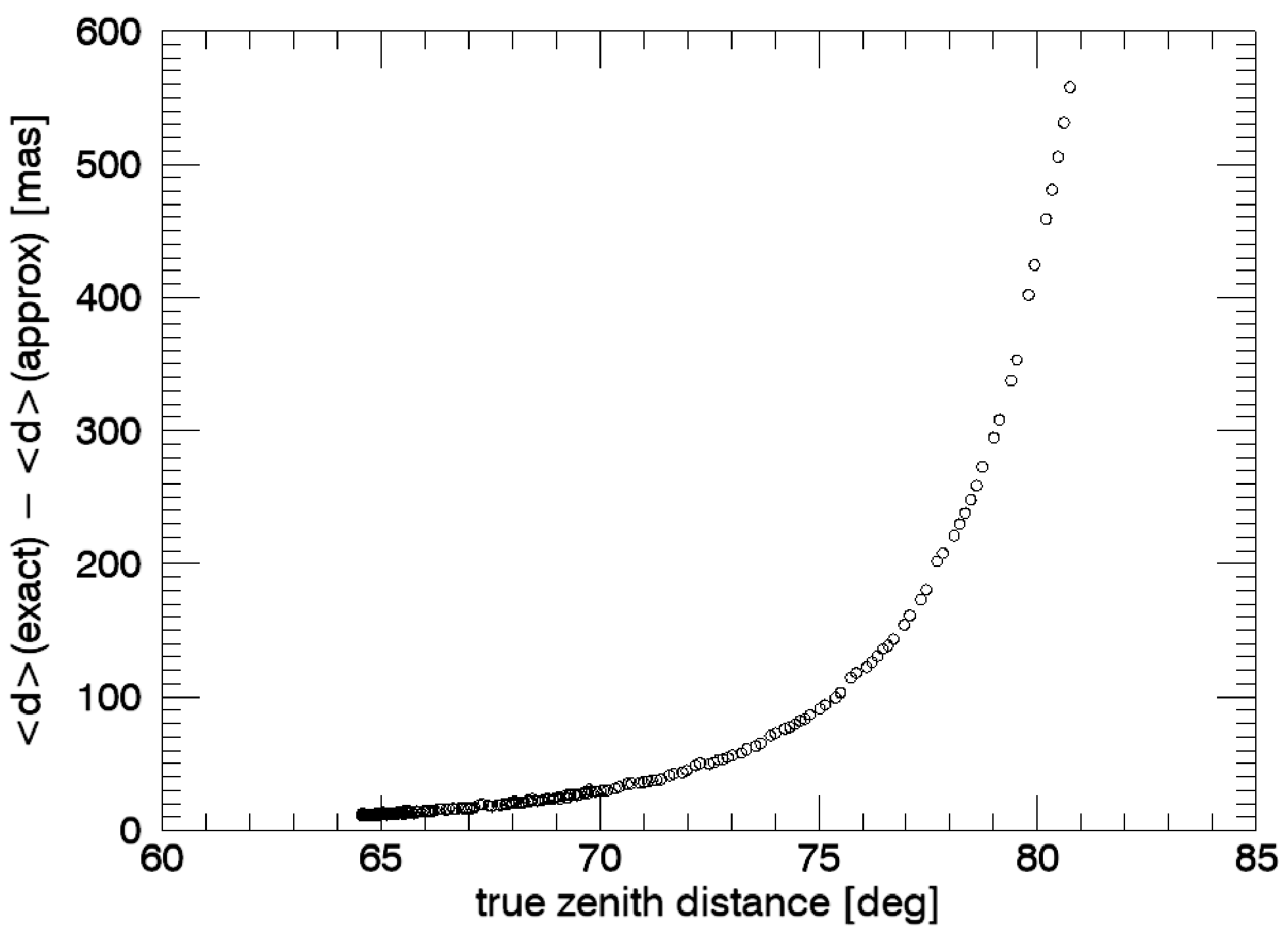}
    \caption{Comparison between mean radius values obtained after refraction correction using exact and approximated formulas. Observation were performed using SODISM-II telescope at Calern Observatory on november 25$^{th}$ 2013.}
              
         \label{Fig:refrac_20131125}
\end{figure}

In order to test the efficiency of the differential refraction estimation, we first compared the refracted solar images  to the theoretical refracted solar shape obtained for the same zenith distances using the direct procedure of Eqs.~(\ref{eq:limb})-(\ref{eq:dist}). The input solar radius in Eq.~(\ref{eq:limb}) is taken equal to the mean value of the observed radii.
Figure~\ref{Fig:refrac_sodism2} shows the good agreement for two sample images recorded at  $z = 64.6\degr$ and $79.5\degr$ on november $25^{th}$ 2013.  For this particular day, figure~\ref{Fig:refrac_20131125} shows, as a function of the zenith distance, the observed difference between the mean corrected solar radius obtained by integration of the exact formula Eq.~(\ref{eq:dist}) and the one obtained using the approximation  Eq.~(\ref{eq:meanell}) applied to the mean of the observed radii. As expected, we recover the curve of figure~\ref{Fig:refrac2diam2}  in the range of the zenith distances covered by the observations. It can be seen that beyond $70\degr$ of zenith distance, the use of the exact formulation is in principle necessary if we want to maintain the correction bias less than 50~mas.

\begin{figure}
  \centering
  \includegraphics[width=.38\textwidth,angle=-90]{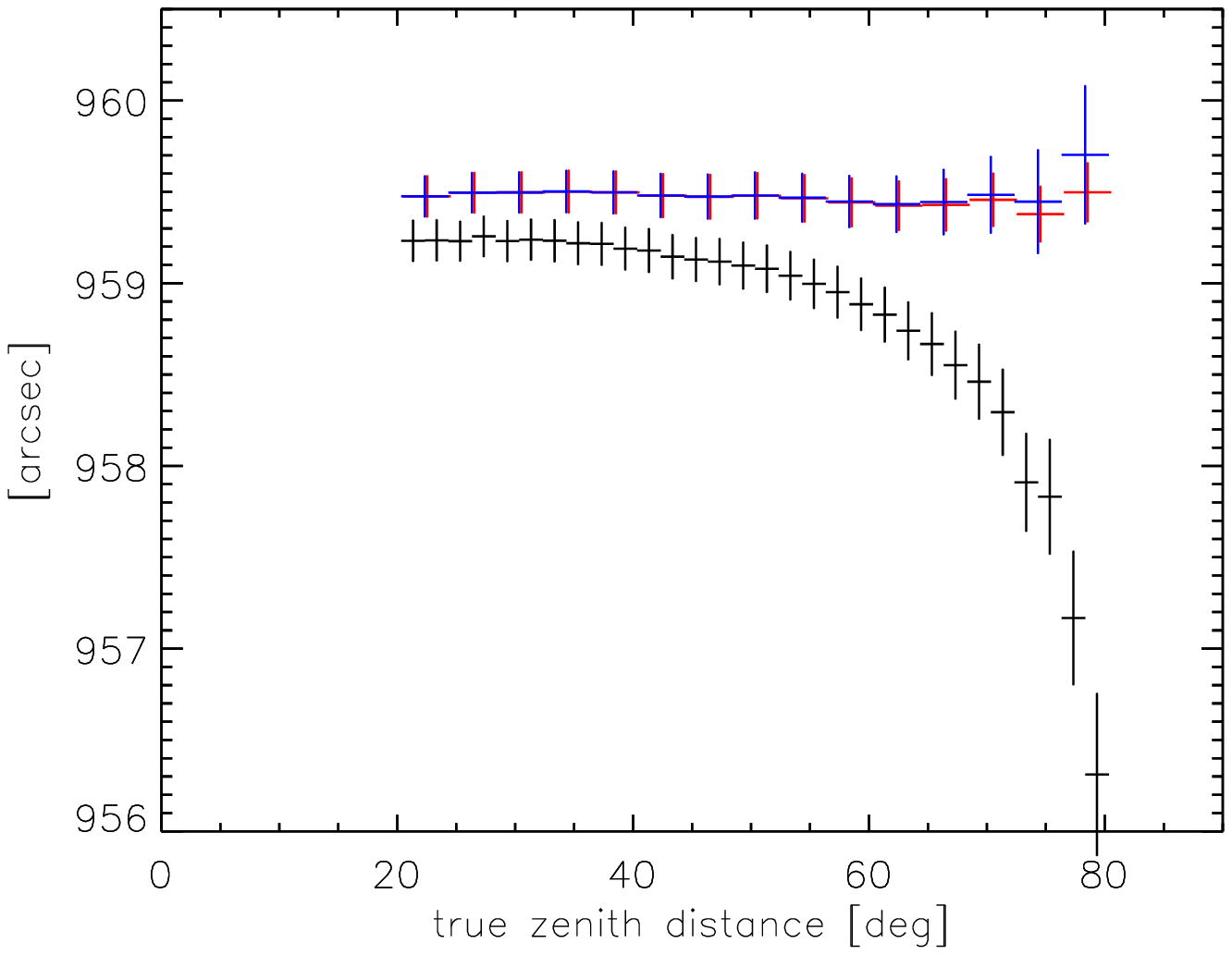}

  \centering
   \includegraphics[width=.38\textwidth,angle=-90]{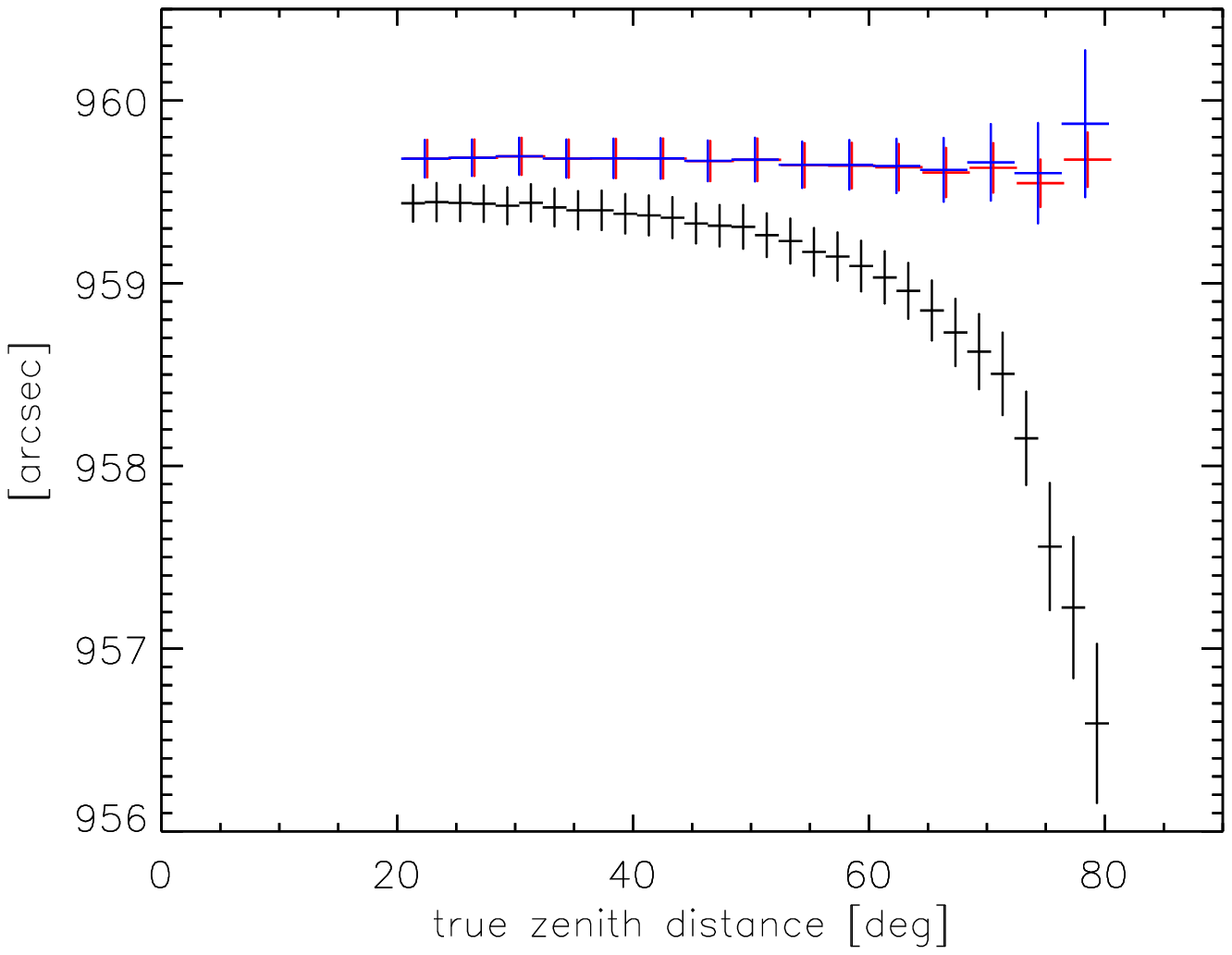}
   \caption{Observed solar radii at 782.2~nm (top) and 1025.0~nm (bottom) during the period 2011-2016 as a function of the true zenith distance. The raw measurements are in black. The blue crosses give the  measurements corrected for  differential refraction by applying  approximate formula Eq~(\ref{eq:meanell}) on the mean radius of each image. The red crosses give the measurements corrected for differential refraction by applying the formulae Eqs.~(\ref{eq:inverse}) on each individual radius of each image. The red crosses have been artificially horizontaly shifted by $0.2\degr$ for clarity.    }
\label{Fig:782-1025}
\end{figure}

We analysed the full set of images following the two approaches. In the first case we compute a mean radius for each image and then apply the approximate mean correction Eq.~(\ref{eq:meanell}). In the second case, we apply the full inverse procedure  Eqs.~(\ref{eq:inverse}) to each individual radii of each image. The results are then grouped  by class of $4\degr$ of zenith distances and a robust estimate of the mean and of the standard deviation is made for each class of zenith distances. On figure~\ref{Fig:782-1025} the raw measurements are shown by the black crosses and the two corrections are shown in blue and red respectively. For the raw measurements we took  zenith distance intervals of $2\degr$ only because of the strong variation at high zenith distance. We see first that the second approach leads to better results for the highest zenith distances covered. The corrected value is in better agreement with the value obtained at low zenith distances. This shows that beyond $75\degr$ we have reached the limit of validity of the approximate formula.  The standard deviation is also significatively lowered. This is a direct consequence of the fact that in the first approach the mean correction applied to each individual radius leads to overestimated  horizontal radii and underestimated vertical radii.

%%%%%%%%%%%%%%%%%%%%%%
\section{Conclusions}\label{sec:conclu}
\begin{figure}
   \centering
 \includegraphics[width=.31\textwidth,angle=90]{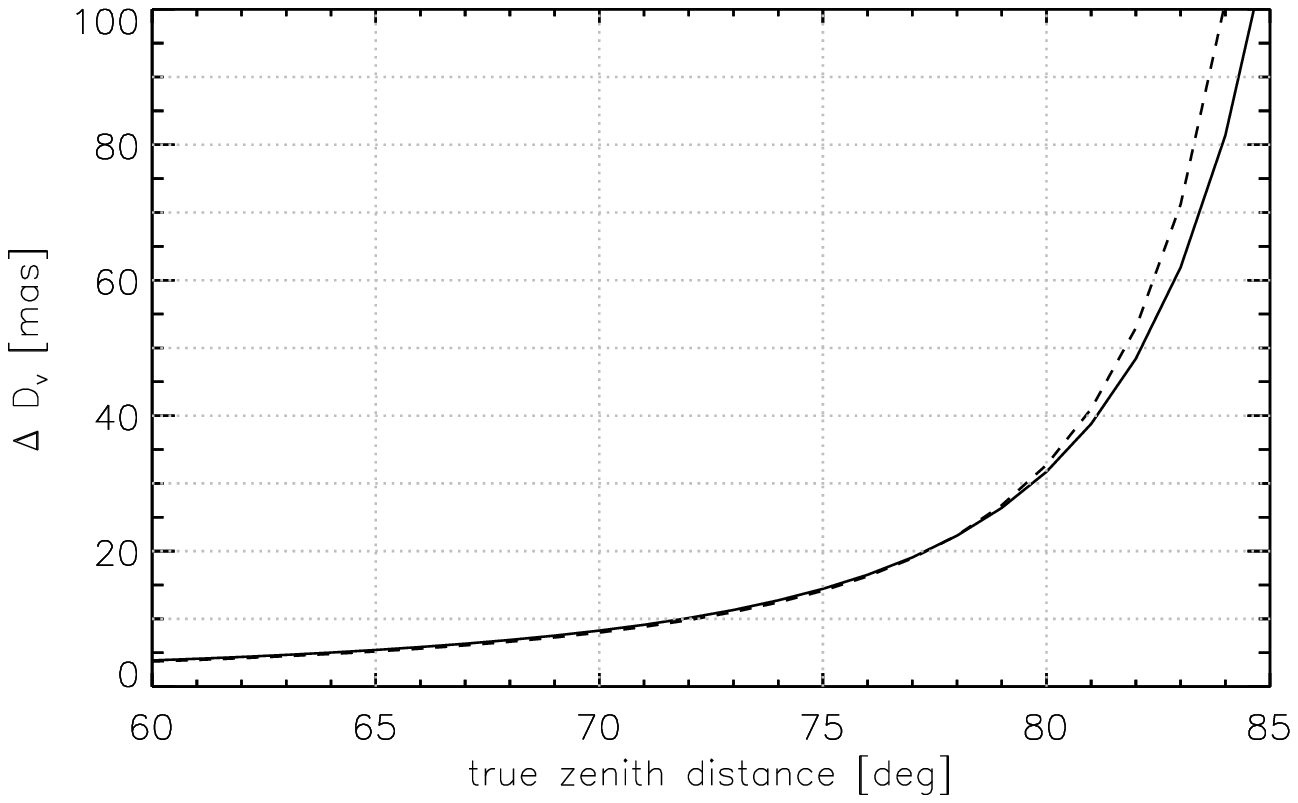}

      \caption{Maximum absolute uncertainties on differential refraction correction for radius measurements in the vertical direction.
 This curves are obtained for  ${\lambda=782.2\ \mbox{nm}}$, ${T=(15 \pm 0.5)\ {\degr}\mbox{C}}$, ${P=(875 \pm 1)\ \mbox{hPa}}$, ${f_h=50\% \pm 5\%}$,  
${dz_{\sun}^t/dt = 650\arcsec\,\mbox{min}^{-1}}$ and 1.43~s of exposure time. 
Full lines give the results from full numerical derivatives calculations while dashed lines are obtained using approximate
formulae Eqs.~(\ref{eq:deriv0})-(\ref{eq:deriv4}). The curves for 1025~nm (not shown) are similar.}
         \label{Fig:refrac2diamfinal}
   \end{figure}

%%%%%%%%%%%%%%%%%%%%%%
The motivation of this work was to address the reliability of the differential refraction correction currently applied to solar astrometric measurements made using full disk imaging. An approximate formula is routinely used for this purpose and for conservative reasons we reject images recorded at zenith distance above $60\degr$. In this work we have analysed about 8000 images recorded above $60\degr$ over a period of five years and we show that the approximate formula for correcting differential refraction from the mean radius is reliable at least up to $70\degr$. For higher zenith distances a more rigourous correction applied to each individual radius as a fonction of the heliographic angle and using the full computation of the refraction integral for a standard atmosphere is able to produce better results up to $80\degr$. 
We have obtained in Section~\ref{Sec:inverse}, the exact formulae that can be used to correct solar radius measurements at any heliographic angle
 and any zenith distance from the effect of astronomical refraction for a given atmospheric model. 
 Absolute uncertainties on these corrections are also derived that 
allows us to fix the maximum zenith distance one should observe depending on the needed metrologic accuracy. Figure~\ref{Fig:refrac2diamfinal} shows the maximum total absolute 
uncertainty obtained on the solar radius assuming that the vertical radii 
have been observed at different zenith distances.  Because we took the maximum value for $\mathrm{d}z_{\sun}^t/\mathrm{d}t$, this curves
represent only upper limits, the actual value of $\mathrm{d}z_{\sun}^t/\mathrm{d}t$ should be use for each measurement.
 From this, one can deduce that observing below $70\degr$, $75\degr$ or $80\degr$ of zenith distances will keep the absolute uncertainties 
on refraction corrections below 10, 20 and 50~mas respectively. The comparison
between numerical derivatives (full lines) and the use of approximate formulae Eqs.~(\ref{eq:deriv0})-(\ref{eq:deriv4}) (dashed lines) shows that, 
even if the approximate formulae should not be used above $70\degr$ for correcting 
the measurements (c.f. Fig.~\ref{Fig:refrac2diam2}), they can be used at least up to ${z_{\sun}^t=80\degr}$ for estimating the uncertainties. 

In summary, the process  that we suggest to correct ground based radii measurements from refraction for true zenith distances up to $80\degr$ is as follow.
Inputs are: the measurements $d(\phi)$ and eventually their associated errors $\delta d(\phi)$ where $\phi$
is an arbitrary angle defined on the solar image; the time of image record and the exposure time $\Delta t$;
weather records ($P$, $T$, $f_h$) and their associated uncertainties ($\Delta T$, $\Delta P$ and $\Delta f_h$); 
the wavelength ($\lambda$) and observer's geodetic coordinates ($\varphi$, $h$).
One can then successively:
\begin{itemize}
\item find the direction of the zenith on the image and associate each angle $\phi$ to its corresponding angle $\psi$ (c.f. Fig.~\ref{Fig:Dessin}).
Depending on the instrumental setup, this may require the computation of the parallactic angle from ephemeris,
\item determine $z_{\sun}^t$ and $\mathrm{d}z_{\sun}^t/\mathrm{d}t$ from ephemeris at the time of image record,
\item calculate $R_{\sun}$ using Eqs.~(\ref{eq:inverse}) and full numerical integration for the refraction function $R(z,\lambda,P,T,f_h,h,\varphi)$,
\item estimate $\Delta d(\psi)$ from Eqs.~(\ref{eq:deriv0})-(\ref{eq:deriv4}) and the knowledge of $\Delta T$, $\Delta P$, $\Delta f_h$, $\Delta t$ and $\mathrm{d}z_{\sun}^t/\mathrm{d}t$,
\item estimate $\Delta R_{\sun}$ from:
\end{itemize}
\begin{equation}
\Delta R_{\sun}=R_{\sun} { {\sqrt{\Delta d(\psi)^2+\delta d(\psi)^2}} \over {d(\psi)} }.
\end{equation}
For zenith distances lower than $70\degr$ full numerical integration can be replaced by Eq.~(\ref{eq:Danjon}) in order to evaluate the refraction function (c.f. Fig.~\ref{Fig:compabs}).
 In both cases Ciddor \citeyearpar{Ciddor1996} equations should be used for computing air refractivity at observer position. 
The corresponding codes are available from the authors upon request.
    
It is important to keep in mind that, at all zenith distances, other phenomena such as extinction or optical 
turbulence must be taken into account for ground based solar metrology. 
We know that they will dominate refraction effects at low zenith distances. 
Close to the horizon extinction is proportional
to refraction (Laplace's extinction theorem) and effects of optical turbulence (e.g. Ikhlef et al. \citeyearpar{Ikhlef2016MNRAS} and reference therein)
 will become increasingly important knowing that the Fried parameter
varies as sec(z)$^{-0.6}$.
It is interesting however to know that for any zenith distance up to $80\degr$ refraction can be reliably corrected and uncertainties on this correction estimated.
After these correction are applied, all  other phenomena impacting  metrologic measurements can therefore be investigated without fearing contamination by astronomical refraction even at high zenith distances.
The mean radius correction presented here (c.f. Fig.~\ref{Fig:refrac2diam}) as well as mean turbulence corrections have been applied to correct the first PICARD-SOL measurements \citep{Meftah2014A&A, Meftah2018}. 
The corrections that can be applied  individually for each heliographic angles should be used in future work in order to disentangle the different effects. In some cases the best seeing conditions are obtained early in the morning when the Sun is still low. This work shows that in such cases the uncertainty associated to refraction correction will be higher but we still can have a good confidence on its magnitude when computed using the rigourous approach instead of the approximate formula. Keeping images recorded between $60\degr$ and $70\degr$ of zenith distances when the seeing conditions are good also potentially provides a way to increase the measurement statistic over the winter periods. 

Finally we note that we have considered only the radial symmetric-component of refraction also called pure or normal refraction. There also exists an asymmetric component 
known as anomalous refraction (e.g. Teleki \citeyear{Teleki1979})  resulting from the tilted atmospheric layers. Anomalous refraction may depend not only on zenith distance but also on azimuth and 
it can lead to seasonal or high frequency effects (see e.g. Hirt \citeyear{Hirt2006} and references therein). The amplitude of such effect has however been found to be lower 
than $0.2\arcsec$ for local effects and one order of magnitude less for regional effects that may originate higher in the atmosphere (e.g. Hu \citeyear{Hu1991}. Moreover it has been shown that anomalous refraction is spatially
coherent at scales of at least $2\degr$ \citep{Pier2003} and it has been established from dedicated observations that its main source is  confined in the layer immediately above ground level (less than 60~m, see \cite{Taylor2013}). It is therefore difficult to believe that differential effects of anomalous refraction and especially 
the one that may be triggered in the 
Upper Troposphere - Lower Stratosphere (UTLS) interface (c.f. \cite{Badache-Damiani2007}) 
could lead to significant bias on solar astrometric measurements relying on direct solar disk imaging. 
 
%%%%%%%%%%%%%%%%%%%%%%%%
\section*{Acknowledgements}
%%%%%%%%%%%%%%%%%%%%%%%%
    This work utilizes data obtained by the PICARD-SOL instruments which are operated by the Observatoire de la C\^ote d{\textquoteright}Azur (OCA) and Laboratoire Atmosph\`eres, Milieux, Observations Spatiales (LATMOS) with the support of the Centre National d{\textquoteright}Etudes Spatiales (CNES) and Programme National Soleil-Terre (PNST).  We thank  P.~Exertier and J.~Paris for useful discussions on the precise geodetic coordinates of Calern station, S. Y. van der Werf (Univ. of Groningen) for providing his ray tracing code also used to check our results and K. Reardon (INAF) for making available his IDL codes for computing refractivity from \cite{Ciddor1996} equations.
 We thank the referee for his valuable remarks and questions that allowed us to improve the quality of the paper. 
This paper is dedicated  to the memory of Francis Laclare who died in 2014 after initiating this work.

\bibliographystyle{mnras}
\bibliography{Bibliography_refraction}

%%%%%%%%%%%%%%%%%%%%%%%%%%%%%%%%%%%%%%%%%%%%%%%%%%%%%%%%%%%%%%%
\appendix 
\section{Note on the radius of curvature at Calern observatory}
\label{App:curvature}
%%%%%%%%%%%%%%%%%%%%%%%%%%%%%%%%%%%%%%%%%%%%%%%%%%%%%%%%%%%%%%%
According to  the  WGS84 reference ellipsoid, the Earth's equatorial and polar radii are 
given respectively by ${a=6378.137\ \mbox{km}}$ and ${b = 6356.752\ \mbox{km}}$.  The 
curvature in the (north-south) meridian and at the geodetic latitude of Calern solar astrometric instruments
${\varphi= 43\degr 45\arcmin 7\arcsec}$ is then given by:
\begin{equation}
r_c^0={(ab)^2 \over {\left(a^2\cos^2(\varphi)+b^2\sin^2(\varphi) \right)^{3/2}}}=6365.985 \ \mathrm{km}
\end{equation}
One could also consider the mean radius of curvature calculated for Calern. From the curvature 
in the prime vertical (normal to the meridian):
\begin{equation}
r_c^{90}={a^2 \over {\sqrt{a^2\cos^2(\varphi)+b^2\sin^2(\varphi)} }}=6388.371 \ \mathrm{km}
\end{equation}
we can deduce the radius of curvature for any azimuth angle $A$ by:
\begin{equation}
 r_c^A={1 \over {{\cos^2(A)\over r_c^0}+{\sin^2(A)\over r_c^{90}}  } }
\end{equation}
from which we can deduce the mean radius of curvature averaging over all directions, by:
\begin{equation}
<\!\!r_c\!\!>=\sqrt{r_c^0 r_c^{90}}= {a^2b \over {a^2\cos^2(\varphi)+b^2\sin^2(\varphi) }}=6377.168 \ \mathrm{km}
\end{equation}
If, instead of the radius of curvature, one considers the distance from geocenter, we have:
\begin{equation}
R=\sqrt{ {{a^4\cos^2(\varphi)+b^4\sin^2(\varphi)} }\over {{a^2\cos^2(\varphi)+b^2\sin^2(\varphi)} }}=6367.955 \ \mathrm{km}
\end{equation}
One should add to these values the elevation of the observer above the reference ellipsoid (${h=1.323\ \mbox{km}}$ for Calern observatory). If we consider that, on average, we observe the sun closer 
to the north-south direction than east-west direction we can take:
\begin{equation}
r_c=r_c^0+h=6367.308 \ \mathrm{km}
\end{equation}
which is very close to the value used by \cite{Chollet1981PhD}.

Finally we note that, for ephemeris calculations, the geodedic latitude should be corrected for the local gravimetric
deflection. For Calern solar astrometric instruments this lead to an astronomic latitude ${\varphi_{\mathrm{ast}}= 43\degr 44\arcmin 53\arcsec}$
which is also compatible within $1\arcsec$ with the direct measurements made using a full entry pupil astrolabe on 
the same site. Similarly, we note that taking into account the local undulation with respect to the reference ellipsoid 
leads to a height above sea level of ${h_{\mathrm{sl}}=1.271\ \mbox{km}}$ for Calern solar astrometric station.

%%%%%%%%%%%%%%%%%%%%%%%%%%%%%%%%%%%%%%%%%%%%%%%%%%%%%%%%%%%%%%%%%%%%%%%%%%%%%%%%%%%%%%%%%%%
\section{Note on the corrections applied to mercurial barometer reading}\label{App:baro}
%%%%%%%%%%%%%%%%%%%%%%%%%%%%%%%%%%%%%%%%%%%%%%%%%%%%%%%%%%%%%%%%%%%%%%%%%%%%%%%%%%%%%%%%%%%
The two corrections (for gravity and barometer temperature) can be written as multiplicative factors (e.g. Princo \citeyearpar{Princo}):
\begin{equation}\label{eq:manuel}
P=H\left({1+L\ \theta}\over{1+M\ \theta}\right){g \over g_0}
\end{equation}
where $P$ is the corrected atmospheric pressure, $H$ is the barometer reading, ${M=1.818\, 10^{-4}\ \mbox{K}^{-1}}$ is the coefficient of volume thermal expansion of mercury,
and ${L=1.84\,10^{-5}\ \mbox{K}^{-1}}$ is the coefficient of linear thermal expansion of brass.
According to the 1967 reference system formula (Helmert's equation), we have:
\begin{equation} \label{eq:helmert}
g=g_{45}\left(1-a\cos(2\varphi)-b\cos^2(2\varphi)\right)
\end{equation}
where ${g_{45}=9.8061999\ \mbox{ms}^{-2}}$ is the gravity acceleration at mid latitude, ${a = 2.64\,10^{-3}}$
and ${b = 1.96\,10^{-6}}$.
This can be corrected from the so-called Free Air Correction (FAC)  which accounts for the fact that gravity 
decreases with height above sea level (${C_{\mathrm{FAC}} = -3.086\, 10^{-6}\ \mbox{s}^{-2}}$), 
itself corrected in order to take into account the increasing gravity due to the extra mass  assumed for a flat terrain
 (Bouger correction, ${C_{\mathrm{B}} = 4.2\, 10^{-10}\ \mbox{m}^3\,\mbox{s}^{-2}\,\mbox{kg}^{-1}}$). 
 For a mean rock density of ${\rho_r = 2.67\, 10^3\ \mbox{kg}\,\mbox{m}^{-3}}$ this leads to:
\begin{equation}
C_g=(C_{\mathrm{FAC}}+\rho_r C_{\mathrm{B}})=-1.96\,10^{-6}\ \mbox{s}^{-2}
\end{equation}
Close to $45\degr$ of latitude, the second term of Eq.~(\ref{eq:helmert}) can be neglected and, 
if we note ${\epsilon = 1-g_{45}/g_0 = 4.6\,10^{-5}}$, Eq.~(\ref{eq:manuel}) can be approximated by:
\begin{equation}
P=H\left(1-\epsilon\right)\ \big(1-(M\!\!-\!\!L)\ \theta\big) \ \left\{1-a\cos(2\varphi) +{C_g\over g_{45}}\  h\right\}
\end{equation}
Neglecting second order terms leads to Eq.~(\ref{eq:baro}).

We note  that absolute gravity measurements have now 
been made at Calern geodetic observatory leading to ${g=(980215549.2 \pm 12.6)\,10^{-8}\ \mbox{m}\,\mbox{s}^{-2}}$ \citep{Nicolas2006}.
 This shows that the relative 
error on the correction
$g/g_0$  discussed  above and previously used for refraction 
calculations was less than $5\,10^{-5}$.  One could however now
directly use Eq.~(\ref{eq:manuel}) with the measured value of local gravity.

\label{lastpage}

\end{document}